\documentclass{aa}  

\usepackage{graphicx}
\usepackage{txfonts}
\usepackage{lipsum}
\usepackage{subcaption}         % necessary for continued figures, example in section 3
\usepackage{lscape}             % to rotate a single page table, example in appendix.
\usepackage{placeins}           % useful with \FloatBarrier, to keep 
\begin{document}

   \title{Plasma frequency waves in Earth's electron foreshock}

   \author{Daniel B. Graham \inst{1}
        \and Yuri V. Khotyaintsev\inst{1}
        \and Mats Andr\'{e}\inst{1}
        \and Andris Vaivads\inst{2,3}
        \and Iver H. Cairns\inst{4}
        }

   \institute{Swedish Institute of Space Physics, Uppsala, Sweden.\\
             \email{dgraham@irfu.se}
            \and Division of Space and Plasma Physics, KTH Royal Institute of Technology, Stockholm, Sweden. 
            \and Ventspils University of Applied Sciences, Ventspils, Latvia.
            \and Department of Physics, University of Sydney, Sydney, New South Wales, Australia. }

   \date{Received \today}
 
  \abstract
   {At Earth's quasi-perpendicular bow shock, electrons can be reflected and accelerated to high velocities, forming suprathermal beams. These beams excite Langmuir and beam-mode waves, which can then be converted to radio waves via linear or nonlinear processes.}
   {We aim to understand the properties and evolution of Langmuir waves excited in the electron foreshock region using the Magnetospheric Multiscale (MMS) mission.} 
   {We use fields and particle data from the four MMS spacecraft to investigate the properties and occurrence of Langmuir/Z-mode waves in Earth's electron foreshock. MMS provides extended high-resolution snapshots of the three-dimensional electric field, enabling detailed analysis of wave properties. Probability distribution functions of the electric field are used to investigate the evolution of the waves and the role of density fluctuations.}
   {Distinct spectral peaks near the electron plasma frequency are often observed, suggestive of simultaneous observations of beam-mode and Langmuir waves, as well as nonlinear electrostatic decay of Langmuir waves or reflection off density gradients. In addition, the electric fields often have large perpendicular components, consistent with low wave number Z-mode waves. The statistical results show that the electric fields are largest near the electron foreshock boundary with the solar wind. Both the parallel and perpendicular components of the electric field exhibit close to log-normal probability distribution functions, consistent with predictions from Stochastic Growth Theory.}
   {These results suggest that small-scale density perturbations in the ambient plasma, in addition to nonlinear three-wave decay, are crucial to the evolution of Langmuir waves and the generation of radio waves at the plasma frequency and second harmonic. These results apply to Langmuir waves in the solar wind, such as in Type II and Type III solar radio burst source regions, where the same density fluctuations are expected and large-amplitude Langmuir waves with similar properties are observed.}

   \keywords{Langmuir waves --
                electron foreshock --
                nonlinear processes
               }

   \maketitle
\nolinenumbers

\section{Introduction}
Langmuir waves are ubiquitous in solar plasmas, such as in solar type II and III radio burst source regions \cite[]{gurnett1976a}, planetary foreshocks \cite[]{filbert1979,anderson1981}, magnetic reconnection regions \cite[]{viberg2013,graham2023}, and auroral regions \cite[]{mcfadden1986,kintner1995}. Langmuir waves are quasi-electrostatic waves with frequencies near the electron plasma frequency and are typically generated by the bump-on-tail instability when fast electron beams are present \cite[]{filbert1979,cairns1987c}. The waves are typically characterized by a narrow spectral range and electric field fluctuations aligned with the background magnetic field. The waves are of particular importance because part of their energy can be converted to electromagnetic waves, leading to radio wave emissions that can be observed remotely, such as in type II and type III radio burst source regions and planetary electron foreshocks. 

Langmuir waves are converted to electromagnetic waves at the electron plasma frequency and its harmonics via the plasma emission mechanism. The basic steps of the processes are: (1) electrons are accelerated to form beams. (2) These beams generate Langmuir waves, which grow to large amplitudes. (3) The Langmuir waves undergo linear or nonlinear processes that convert part of their energy into electromagnetic waves, which can propagate away and be observed remotely. Various processes have been proposed as mechanisms for converting Langmuir waves to electromagnetic radio waves, including linear-mode conversion \cite[]{field1956,kim2007}, electrostatic decay and coalescence \cite[]{cairns1985,cairns1987b,cairns1988}, electromagnetic decay \cite[]{cairns1987a}, and antenna mechanisms \cite[]{papadopoulos1978,malaspina2010a}. 

Observations in the electron foreshock and type III source regions have found evidence that electrostatic decay frequently occurs. The observation of Langmuir waves with two distinct spectral peaks has been proposed as evidence of three-wave electrostatic decay \cite[]{cairns1992a,cairns1995,hospodarsky1995,soucek2005,henri2011,graham2013a}, where beam-driven Langmuir waves decay into counter-propagating Langmuir waves and ion-acoustic waves. Others have argued that the observations of counter-propagating Langmuir waves could be due to reflection off density gradients rather than electrostatic decay \cite[]{krasnoselskikh2011}. 

One of the features of Langmuir waves observed in space plasmas is that they are highly bursty, meaning their amplitude can vary by orders of magnitude over very short time intervals \cite[]{gurnett1977,lin1981,lin1986}. These bursty Langmuir waves have been interpreted in terms of nonlinear processes, such as electrostatic decay \cite[]{cairns1985,lin1986}. However, density fluctuations in the solar wind can strongly affect the evolution of Langmuir waves. \cite{smith1979} proposed that clumping Langmuir waves can be the result of density inhomogeneities rather than nonlinear processes. Density fluctuations modify the refractive index of Langmuir waves, which can cause Langmuir waves to shift wave numbers or reflect off density gradients. 

In \cite{robinson1992a}, Stochastic Growth Theory (SGT) was proposed to explain the highly bursty nature of Langmuir waves. \cite{robinson1992a} proposed that density fluctuations shift the wave number of Langmuir waves in and out of resonance with the electron beam, resulting in regions of growth and damping of the waves, and an overall small net energy transfer to the waves. By applying the central limit theorem for a large number of fluctuations in the wave gain, the probability distribution of the electric field amplitude is predicted to follow a log-normal distribution in the absence of nonlinear processes. Observations of the Langmuir waves in the electron foreshock and type III source regions found that the probability distributions closely matched log-normal distributions \cite[]{robinson1993a,cairns1997b,cairns1999,sigsbee2004a,sigsbee2004b,cairns2026}. Others have argued that the probability distributions of Langmuir waves differ from the SGT prediction and that the Langmuir wave distributions are better modeled by the Pearson system of distribution functions \cite[]{musatenko2007,krasnoselskikh2007,vidojevic2011,vidojevic2014}.

Recent observations of Langmuir waves in Earth's electron foreshock and in type III source regions have found that the waves can often have large electric fields perpendicular to the background magnetic field \cite[]{bale1998b,malaspina2011,graham2013a}. These strong perpendicular fields were interpreted as low-wave number Z-mode waves on the generalized Langmuir/Z-mode dispersion surface. Recent observations from Parker Solar Probe and Solar Orbiter were able to measure the magnetic field fluctuations associated with these waves, confirming that the waves are consistent with Z-mode (also termed slow extraordinary) waves \cite[]{larosa2022,formanek2025}. 

Although Langmuir waves in Earth's electron foreshock have been studied extensively since they were first discovered there, the data from the MMS spacecraft can provide new insights into the properties and evolution of Langmuir waves by providing longer waveform capture intervals and three-dimensional electric field measurements. In this paper, we investigate the properties of Langmuir waves in Earth's electron foreshock using the MMS spacecraft. The outline of this paper is as follows: In section \ref{mmsdata} we present the data used. In section \ref{wavepropssec} we investigate the properties of waves observed in the electron foreshock. Section \ref{stats} provides a statistical overview of the waves in the electron foreshock. In section \ref{fieldstatssec} we investigate the behavior of Langmuir waves using field statistics of Langmuir waveforms and compare the results with SGT. Sections \ref{discussion} and \ref{conclusions} are the discussion and conclusions, respectively. 

\section{Data} \label{mmsdata}
We use data from the MMS spacecraft, specifically electric field ${\bf E}$ data from the Electric field Double Probes (EDP) instruments \cite[]{lindqvist1,ergun3}, magnetic field ${\bf B}$ data from the Fluxgate Magnetometer (FGM) \cite[]{russell2}, and particle data from the Fast Plasma Investigation (FPI) \cite[]{pollock1}. The four MMS spacecraft orbit Earth in a close tetrahedral configuration, with nominal spacecraft separations of a few tens of kilometers. For the time interval considered here, from May 2017 to June 2022, the apogee was $\approx 25 - 30 \, R_E$, where $R_E$ is Earth's radius, meaning that MMS spent significant time in the solar wind and electron foreshock. All data presented in this paper are high-resolution burst mode data. To measure the ${\bf E}$ waveforms of plasma frequency waves, we use the AC coupled hmfe data product, which provides high-resolution snapshots of ${\bf E}$ in three dimensions. This data product captures snapshots of ${\bf E}$ with nominal durations of 2 seconds and a sampling rate of $65.536$~kHz. The hmfe snapshots are acquired when burst mode selections are made by the Scientist in the Loop (SITL) \cite[]{fuselier2016}. Within the burst mode intervals, the snapshots are selected based on electric field fluctuations, with a duty cycle of $25$~\%. This means that the intervals with hmfe snapshots are biased towards times deemed interesting, such as magnetopause crossings, magnetic reconnection events, bow shock crossings, and structures in the solar wind. Nevertheless, a substantial amount of burst data has been recorded in the electron foreshock and in the solar wind. 

The MMS spacecraft have several advantages over previous spacecraft in investigating Langmuir waves. These are: 

(1) The high-frequency waveforms sampled by MMS have much longer durations (nominally 2 seconds) compared with previous missions. For example, the high-resolution waveforms were captured over several tens of milliseconds for Wind \cite[]{bougeret1995}, Solar Terrestrial Relations Observatory (STEREO) \cite[]{bale2008}, and Solar Orbiter \cite[]{maksimovic2020}. 

(2) Three-dimensional electric fields are measured. Three-dimensional electric fields were also measured by STEREO using three monopole antennas \cite[]{bale2008}, while other spacecraft typically measured one or two orthogonal components of the electric field.  

(3) Electron distributions and moments are measured at substantially higher temporal resolutions than previous missions. This can provide new insights into how the electron beam evolves and how background density perturbations affect the growth and evolution of plasma frequency waves. 

The main disadvantage of using MMS data is that the typical Nyquist frequency of the highest-resolution electric field data is $32.768$~kHz, which is typically sufficient to resolve plasma frequency waves in the solar wind. However, for dense solar wind, with an electron number density of $n_e \gtrsim 13.3$~cm$^{-3}$, the plasma frequency can exceed the Nyquist frequency, preventing plasma frequency waves from being observed. 

\section{Wave properties and examples} \label{wavepropssec}
\begin{figure}[htbp!]
\begin{center}
\includegraphics[width=90mm, height=130mm]{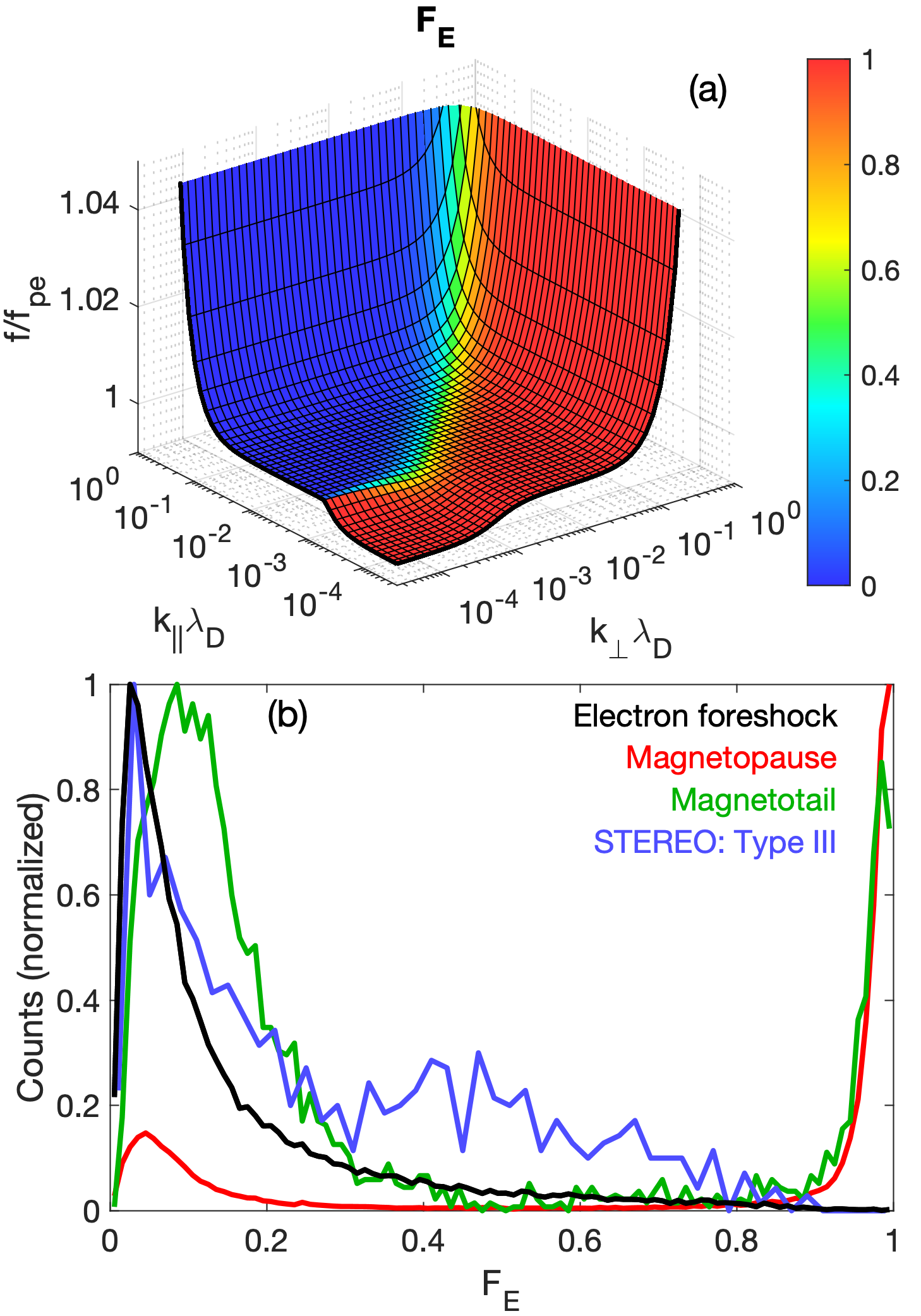}
\caption{Overview of the fraction of perpendicular energy density to total energy density for Langmuir/Z-mode and upper hybrid (UH) waves. (a) Langmuir/Z-mode and UH dispersion surface for electron foreshock conditions. The color shading indicates $F_E$ as a function of $f$ and ${\bf k}$. (b) Histograms for plasma frequency waves with $E_{\mathrm{max}} > 5$~mV~m$^{-1}$ versus $F_E$ in Earth's foreshock (black), near Earth's magnetopause (red), in Earth's magnetotail (green), and in type III solar radio bursts observed by STEREO (blue). }
\label{FEoverview}
\end{center}
\end{figure}

In this section, we provide an overview of the plasma frequency waves observed by MMS. Figure \ref{FEoverview}a shows the dispersion surface of Langmuir waves over a narrow frequency range around the electron plasma frequency $f_{pe}$. Langmuir waves, upper hybrid waves, and Z-mode waves all lie on this same dispersion surface. The dispersion surface is calculated using WHAMP (waves in homogeneous, anisotropic, multi-component plasmas) \cite[]{ronnmark1}. As input, we use a single Maxwellian electron distribution and nominal solar wind conditions at 1 astronomical unit (au): electron number density $n_e = 2$~cm$^{-3}$, electron temperature $T_e = 10$~eV, and $|{\bf B}| = 5$~nT. The color shading indicates $F_E = E_{\perp}^2/(E_{\perp}^2 + E_{\parallel}^2)$, where $E_{\perp}$ and $E_{\parallel}$ are the magnitudes of the electric field perpendicular and parallel to the ambient ${\bf B}$. For Langmuir waves with $k_{\parallel} \gg k_{\perp}$, $F_E \approx 0$, corresponding to $E_{\parallel} \gg E_{\perp}$, where $k_{\parallel}$ and $k_{\perp}$ are wave numbers parallel and perpendicular to ${\bf B}$. At low $k$, the electrostatic Langmuir wave couples to the electromagnetic Z-mode wave \cite[]{andre1985,willes2000}, forming a single dispersion surface. For small $k$, the Z-mode wave is left-hand circularly polarized with $E_{\perp} \gg E_{\parallel}$, corresponding to $F_E \approx 1$. For parallel propagation, the mode is electromagnetic, while for oblique wave-normal angles, the Z-mode has both electrostatic and electromagnetic components. At small $k$, plasma oscillations at $f_{pe}$ are connected to the L-O mode dispersion surface above the dispersion surface shown in Figure \ref{FEoverview}a \cite[]{andre1985,graham2018a}. For UH waves with $k_{\perp} \gg k_{\parallel}$, the waves are quasi-electrostatic with $F_E \approx 1$, corresponding to $E_{\perp} \gg E_{\parallel}$. Intermediate values of $F_E$ occur for oblique waves with comparable $k_{\parallel}$ and $k_{\perp}$. Thus, the value of $F_E$ can be used to distinguish Langmuir waves from UH or Z-mode waves. 

We compile a database of hmfe snapshots containing plasma frequency waves from May 2017 to June 2022 \cite[]{graham2023}. Overall, we identify $110,079$ snapshots with plasma frequency waves, where the maximum electric field $E_{\mathrm{max}}$ exceeds $5$~mV~m$^{-1}$. We classify these snapshots into three different regions, namely, in the solar wind, near the magnetopause, and in the magnetotail. We find 43,664 snapshots in the solar wind, 63,535 snapshots near the magnetopause, and 2,880 in the magnetotail \cite[]{graham2023}. The snapshots identified in the solar wind are almost exclusively observed in the electron foreshock, where the magnetic field is connected to Earth's bow shock, rather than in the pristine solar wind, so we consider these electron foreshock waves. 

Figure \ref{FEoverview}b shows the histograms of $F_E = \sum E_{\perp}^2/(E_{\perp}^2 + E_{\parallel}^2)$ for the plasma frequency waves in Earth's electron foreshock, where $F_E$ is calculated for each snapshot by summing over the entire snapshot. For comparison, we also plot the histograms of $F_E$ for waves near the magnetopause [cf., \cite{graham2018a}] and in the magnetotail \cite[]{graham2023}, and Langmuir/Z-mode waves observed in type III source regions from STEREO observations based on 732 snapshots \cite[]{graham2014b}. Near the magnetopause and in the magnetotail, $F_E \approx 0$ and $F_E \approx 1$, corresponding to Langmuir and upper hybrid waves, respectively \cite[]{graham2018a,graham2023}.

In the electron foreshock, the histogram peaks at $F_E \approx 0.025$ with a median $F_E$ of $0.089$. This means that most of the waves are either field-aligned Langmuir waves or beam-mode waves. Beam modes are electrostatic waves with frequencies above and below the electron plasma frequency \cite[]{fuselier1985}, and can only occur when dense beams are present. There are also some snapshots with intermediate values of $F_E$, indicating significant contributions of $E_{\perp}$ to the total electric field power. For $F_E \gtrsim 0.9$, the number of snapshots is negligible, in contrast to observations near Earth's magnetopause and in the magnetotail, where UH waves with $F_E \gtrsim 0.9$ are common. The lack of snapshots with $F_E \gtrsim 0.9$ indicates that Z-mode or UH waves alone are extremely rare or do not occur in the electron foreshock. 

We find that the distribution of $F_E$ in the electron foreshock also differs from the distribution obtained from type III source regions. In particular, in type III source regions, there is a higher proportion of snapshots with intermediate values of $F_E$. We take $F_E = 0.2$ as a threshold to distinguish between snapshots with small and large contributions of $E_{\perp}$ to the total wave power. For the foreshock waves, we find that $\approx 75 \, \%$ of the snapshots have $F_E < 0.2$, while for type III source regions $\approx 50 \, \%$ of the snapshots had $F_E < 0.2$. Thus, significant $E_{\perp}$ are more likely in type III source regions compared with the electron foreshock. Previous observations of type III source regions found that waves with large $F_E$ were associated with faster electron beams \cite[]{malaspina2011,graham2013a,graham2014b}, which may suggest that, on average, the electron beam speeds in the electron foreshock are slower than those in type III source regions at 1~au. 

\begin{figure*}[htbp!]
\begin{center}
\includegraphics[width=160mm, height=120mm]{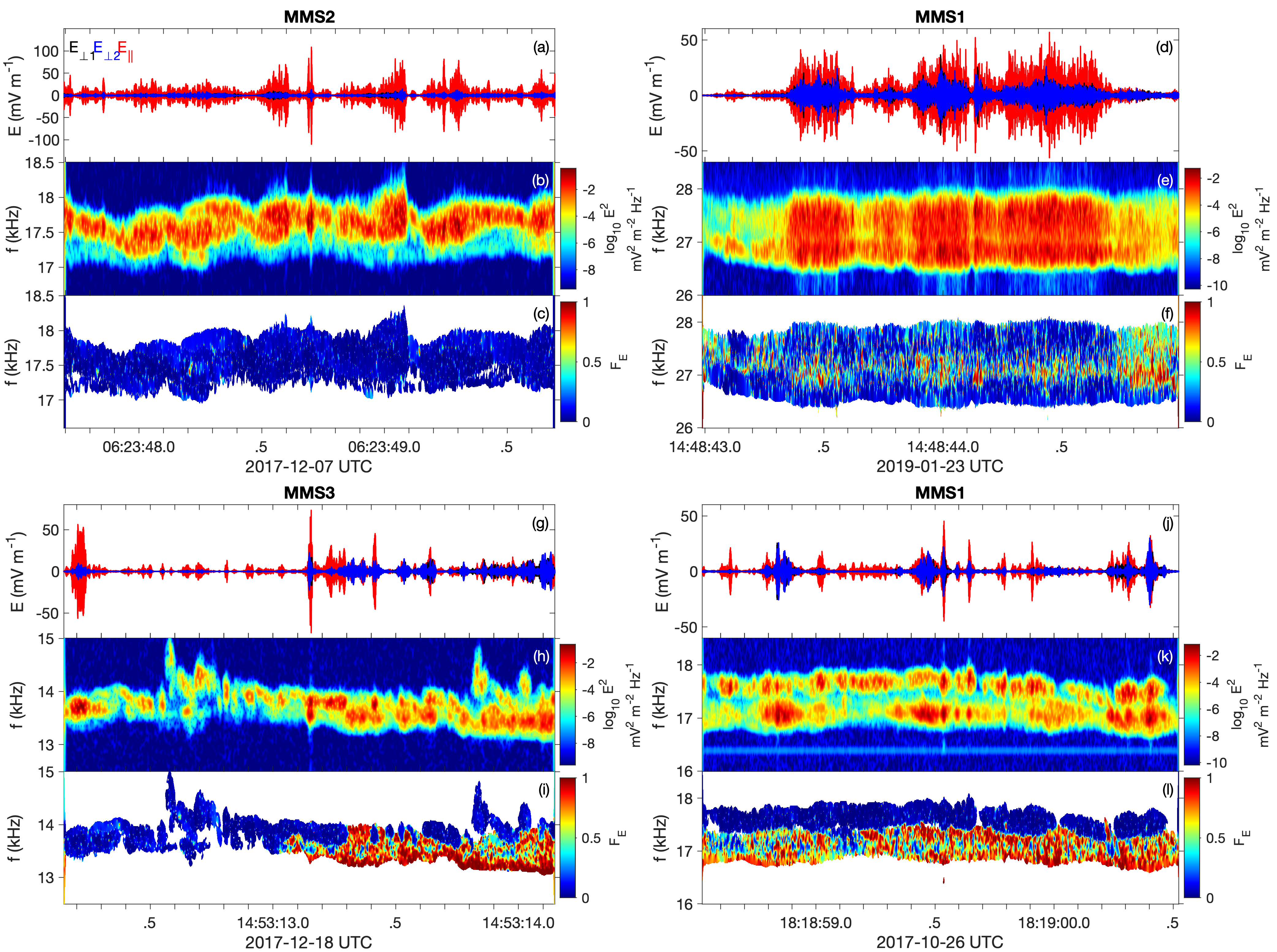}
\caption{Four examples of Langmuir/Z-mode waves in the electron foreshock. (a)--(c) Field-aligned Langmuir wave. (d)--(f) Field-aligned Langmuir waves with two spectral peaks. (g)--(i) Langmuir/Z-mode waves where large perpendicular electric fields are observed. (j)--(l) Langmuir/Z-mode waves with two spectral peaks. (a), (d), (g), and (j) Electric field in field-aligned coordinates $E_{\parallel}$ (red), $E_{\perp1}$ (black), and $E_{\perp2}$ (blue). (b), (e), (h), and (k) Spectrograms of ${\bf E}$. (c), (f), (i), and (l) Spectrograms of $F_E$.}
\label{LZexamples}
\end{center}
\end{figure*}

We now present some examples of Langmuir waveforms observed in the electron foreshock. Figure \ref{LZexamples} shows four examples of the types of Langmuir waves observed by MMS. In each case, we plot the electric field ${\bf E}$ in field-aligned coordinates, where $E_{\parallel}$ is parallel to ${\bf B}$, and $E_{\perp1}$ and $E_{\perp2}$ are perpendicular to ${\bf B}$, the frequency-time power spectrum of ${\bf E}$, and the frequency-time spectrogram of $F_E$. Figures \ref{LZexamples}a--\ref{LZexamples}c show an example of Langmuir waves characterized by $E_{\parallel} \gg E_{\perp}$. Throughout the interval, $F_E$ remains close to $0$ (Figure \ref{LZexamples}c). Over this period of time, we observe complex changes in the wave frequency with time (Figure \ref{LZexamples}b). Similarly, the amplitude and wave power are highly variable throughout the interval. The Langmuir waveform consists of localized clumps, which can have durations as short as a few milliseconds (ms). Figures \ref{LZexamples}d--\ref{LZexamples}f show a Langmuir wave where ${\bf E}$ is primarily aligned with ${\bf B}$. These Langmuir waves are characterized by two distinct spectral peaks separated by $\approx 700$~Hz throughout most of the snapshot, and result from Doppler shift in the solar wind of waves with distinct ${\bf k}$. The two spectral peaks result in rapid fluctuations in amplitude, as seen in Figure \ref{LZexamples}d, due to beating between the waves with distinct frequencies. In addition, slower variations in the amplitude are also observed. Langmuir waves with two spectral peaks are interpreted as evidence of electrostatic decay \cite[]{cairns1992a,cairns1995,hospodarsky1995,graham2013a} or Langmuir waves reflected off density gradients \cite[]{krasnoselskikh2011}. 

The examples in Figures \ref{LZexamples}g--\ref{LZexamples}i and \ref{LZexamples}j--\ref{LZexamples}l exhibit strong $E_{\perp}$ and large $F_E$. The waveform in Figures \ref{LZexamples}g--\ref{LZexamples}i is characterized by bursty wave activity. In the first half of the interval $F_E \approx 0$, corresponding to field-aligned Langmuir waves, while in the second half, most of the waves are characterized by $F_E \sim 1$, with some patches where $F_E \approx 0$ (Figure \ref{LZexamples}i). At some times, waves are observed simultaneously with $F_E \approx 0$ and $F_E \sim 1$ at distinct frequencies. When the $F_E \approx 0$ and $F_E \sim 1$ waves occurred simultaneously, the $F_E \sim 1$ waves often occurred at lower frequencies, consistent with low-$k$ Z-mode waves rather than UH waves. We note that for Langmuir waves propagating anti-sunward, the wave frequencies are increased due to Doppler shift, while the frequencies are decreased for sunward propagating waves. The final example, Figures \ref{LZexamples}j--\ref{LZexamples}l, is characterized by two distinct spectral peaks with $\Delta f = 500$~Hz. In this example, the lower spectral peak is characterized by large $F_E$, while the upper spectral peak is characterized by $F_E \approx 0$. Figure \ref{LZexamples}k shows that the powers in the upper and lower frequency bands are highly variable and are not strongly correlated with each other. This case is consistent with Langmuir and Z-mode waves occurring simultaneously. We note that for fast electron beams, Langmuir waves can undergo three-wave decay to Z-mode waves \cite[]{graham2013a,layden2013,cairns2018}. 

In these two examples, the combination of Langmuir waves with $F_{E} \approx 0$ and Z-mode waves with $F_E \sim 1$ results in snapshots with intermediate values of $F_E$ in Figure \ref{FEoverview}b. We conclude that the observed waves are Langmuir/Z-mode waves. In most cases, the waves are characterized by $F_{E} \approx 0$, corresponding to field-aligned electrostatic Langmuir waves. However, strong perpendicular electric fields are also commonly observed, consistent with low wave-number Z-mode waves. The examples show that the waveforms are highly variable in amplitude and vary with frequency over time. 

\begin{figure*}[htbp!]
\begin{center}
\includegraphics[width=160mm, height=60mm]{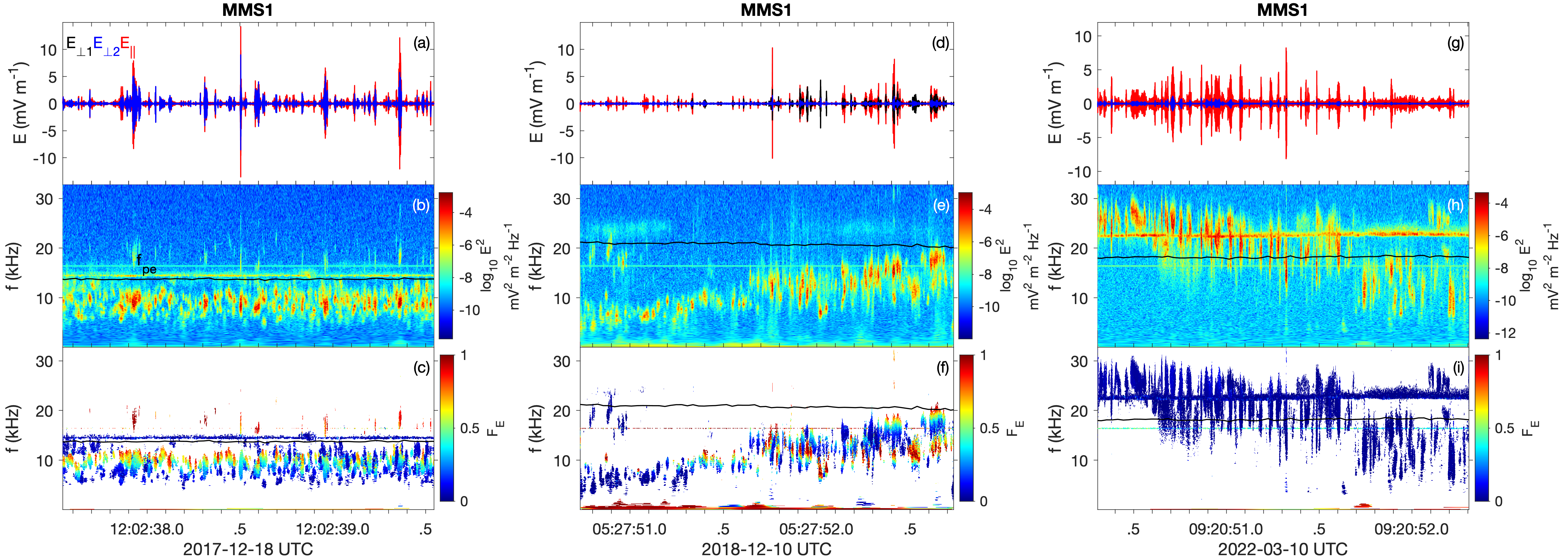}
\caption{Three examples of beam-mode-like waves in the electron foreshock. (a)--(c) Beam-mode wave below $f_{pe}$ and Langmuir wave. (d)--(f) Beam mode wave below $f_{pe}$ without Langmuir wave emission. (g)--(i) Beam-mode wave above and below $f_{pe}$ and Langmuir waves. (a), (d), and (g) Electric field in field-aligned coordinates $E_{\parallel}$ (red), $E_{\perp1}$ (black), and $E_{\perp2}$ (blue). (b), (e), and (h) Spectrograms of ${\bf E}$. (c), (f), and (i) Spectrograms of $F_E$. The black lines in panels (b)--(c), (e)--(f), and (h)--(i) are $f_{pe}$ calculated from $n_e$ obtained from the FPI electron density.}
\label{beamexamples}
\end{center}
\end{figure*}

In addition to Langmuir/Z-mode waves, beam-mode waves also develop in the electron foreshock. Beam-mode waves are typically more broadband in frequency than Langmuir waves and can have wave power both above and below $f_{pe}$ \cite[]{fuselier1985,cairns1989,soucek2019}. These authors showed that beam-mode waves are typically produced for electron beams with $v_b \lesssim 3 v_e$, where $v_b$ is the beam speed and $v_e$ is the electron thermal speed. As examples of the types of beam-mode wave snapshots seen by MMS, Figure \ref{beamexamples} shows three hmfe snapshots of beam-mode waves. The panels are in the same format as Figure \ref{LZexamples}. Figures \ref{beamexamples}a--\ref{beamexamples}c show beam-mode waves with frequencies below $f_{pe}$. Figure \ref{beamexamples}a shows that $E_{\parallel}$ is the dominant component of ${\bf E}$, although strong $E_{\perp}$ develops, resulting in ${\bf E}$ being oblique to ${\bf B}$. The waveform is characterized by highly localized increases in ${\bf E}$. Figure \ref{beamexamples}b shows that the beam-mode wave power is below $f_{pe}$ and is much broader in frequency than Langmuir waves. Low-amplitude Langmuir waves with $f \approx f_{pe}$ are observed in Figure \ref{beamexamples}b. 

Figures \ref{beamexamples}d--\ref{beamexamples}f show a beam-mode wave example with $f < f_{pe}$, with $f$ increasing with time. In this case, no Langmuir waves are observed over the interval. The waveform of ${\bf E}$ is oblique to ${\bf B}$ and is characterized by localized increases in ${\bf E}$. Figures \ref{beamexamples}g--\ref{beamexamples}i show a third example of beam-mode waves. In this case, the beam-mode waves decrease in frequency from above $f_{pe}$ to below $f_{pe}$ over the snapshot interval. In addition, we observe strong Langmuir waves near $f_{pe}$ at a relatively constant frequency. For both the Langmuir waves and the beam-mode waves, $F_E \sim 0$, corresponding to $E_{\parallel} \gg E_{\perp}$. Like the previous examples, the waveform is characterized by highly localized increases in ${\bf E}$. 

These three events illustrate the typical characteristics of the beam-mode waves. Namely, that beam-mode waves are broadband in frequency compared with Langmuir waves, with frequencies below and above $f_{pe}$. The waveforms are characterized by bursty electric field enhancements and can develop simultaneously with Langmuir waves, as well as independently. 

\section{Statistical results} \label{stats}
In this section, we present a statistical overview of the occurrence and properties of the waves near $f_{pe}$ in the electron foreshock. We analyze all hmfe snapshots recorded in the solar wind when the solar conditions are relatively constant, so their locations with respect to the boundary between the electron foreshock and the solar wind can be estimated with minimal uncertainties. 

\subsection{Bow shock model and foreshock coordinates} 
\begin{figure}[htbp!]
\begin{center}
\includegraphics[width=60mm, height=100mm]{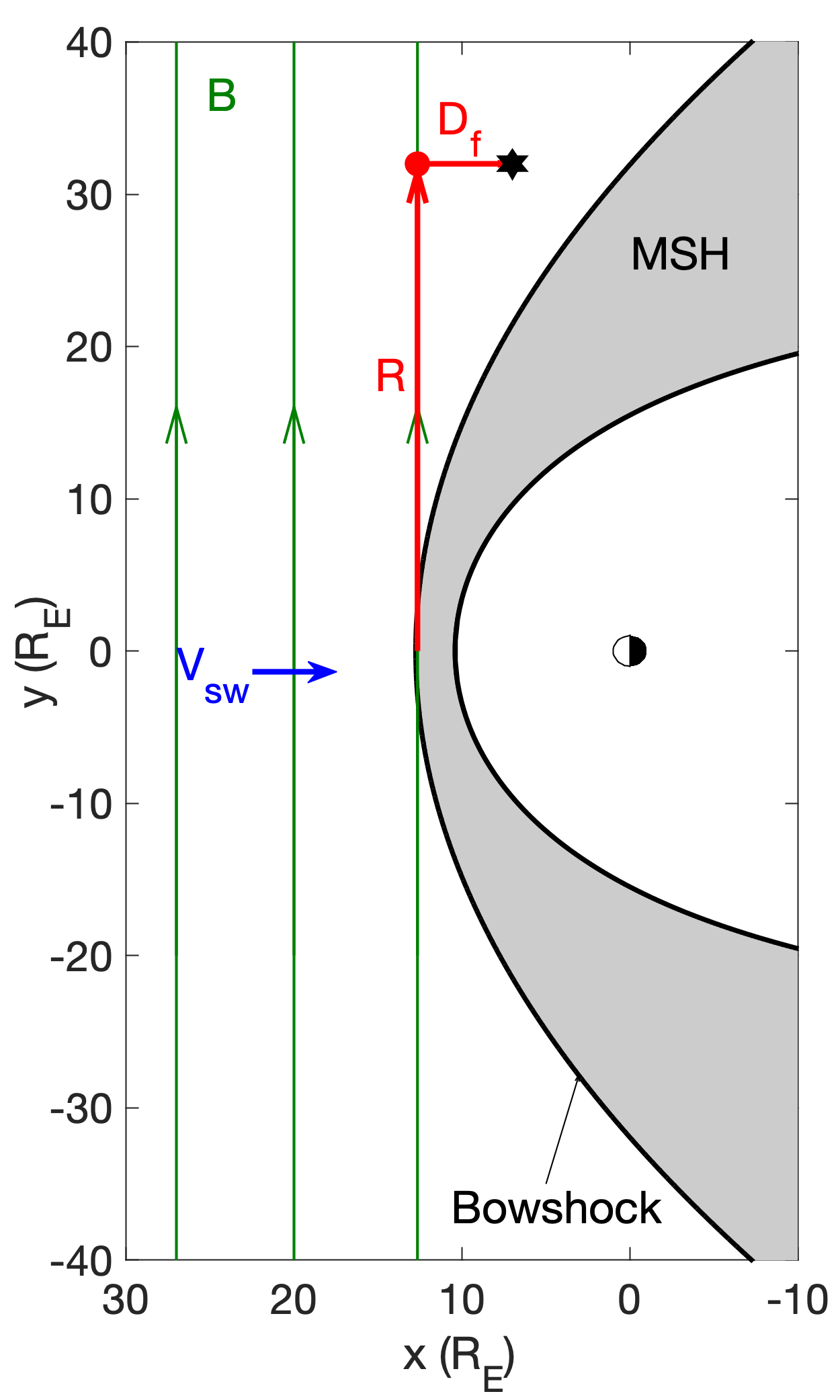}
\caption{Illustration of Earth's bow shock and electron foreshock, and the foreshock coordinate system in the x--y plane in GSE coordinates. Earth is located at (0,0) and the outer and inner black lines show the bow shock and magnetopause, respectively. The gray-shaded region is the magnetosheath (MSH). The green lines and blue arrow represent the solar wind magnetic field and solar wind flow. The red arrows indicate the foreshock coordinates $R$ and $D_f$ for the spacecraft at the location indicated by the star. }
\label{schematic}
\end{center}
\end{figure}

To organize the data and determine where the waves occur, we convert the positions from GSE coordinates, where the hmfe snapshots were taken, to a foreshock coordinate system. In particular, we are interested in the distance $D_f$ downstream of the boundary between the electron foreshock along the solar wind flow, and the distance along ${\bf B}$, which has a tangent with the bow shock, to the spacecraft $R$. Here, $D_f > 0$ corresponds to positions downstream of the tangent ${\bf B}$ in the electron foreshock, while $D_f < 0$ corresponds to solar wind that is not magnetically connected to the bow shock. The procedure to calculate these foreshock coordinates is detailed in \cite{cairns1997a}. Figure \ref{schematic} shows a schematic of Earth's bow shock and the foreshock coordinates. Electrons from the solar wind can be reflected and accelerated at the quasi-perpendicular bow shock. As reflected electrons propagate away from the bow shock along the magnetic field, they are convected downstream by the solar wind. This results in a velocity filter effect, where the fastest electrons are observed at small $D_f$ \cite[]{filbert1979,cairns1987c,cairns1997a}, while slower electrons have more time to convect further downstream, increasing $D_f$. Therefore, electron beam speeds are predicted to decrease as $D_f$ increases. 

To model the bow shock and order the solar wind data, we assume that Earth's bow shock is a paraboloid given by
\begin{equation}
X = a_s - b_s \left( Y^2 + Z^2 \right), 
\label{bowshockmodel1}
\end{equation}
where $a_s$ and $b_s$ are the standoff distance and curvature of the paraboloid, respectively. The two parameters 
are given by \cite[]{kuncic2004,malaspina2009}
\begin{equation}
a_s = 14.2 D_p^{-1/6} R_E,
\label{as}
\end{equation}
\begin{equation}
b_s = a_s \left( 27 + 84 \frac{V_A^2}{V_{sw}^2} \right)^{-2} D_p^{1/3} R_E^{-2}, 
\label{bs}
\end{equation}
where $D_p = n_e m_p V_{sw}^2$ is the solar wind dynamic pressure, $V_A$ is the Alfv\'{e}n speed, $V_{sw}$ is the solar wind speed, $m_p$ is the proton mass, and $R_E$ is Earth's radius. To calculate $a_s$ and $b_s$, we use the local solar wind and electron foreshock parameters measured by MMS. From this foreshock model and the direction of ${\bf B}$, we calculate $D_f$ and $R$, as illustrated in Figure \ref{schematic}.

\subsection{Statistical overview} \label{subsecstats}
To statistically study the plasma frequency waves in the electron foreshock, we investigate all hmfe waveforms in the solar wind. We select all burst mode intervals where MMS was in the solar wind, identified using the Spin-plane Double Probe (SDP) region calibration files. We also only consider hmfe snapshots when the median $f_{pe}$ across the snapshot time, calculated from the FPI electron moments, is below the Nyquist frequency of the ${\bf E}$ waveforms. Furthermore, we only consider hmfe snapshots where the magnetic field does not change too much across the interval, so the foreshock coordinates can be reasonably calculated. Specifically, we only consider hmfe intervals when the direction of ${\bf B}$ changes by less than $10^\circ$ and the magnitude of ${\bf B}$ does not change by more than $20~\%$. 
%Likewise, we only include snapshots where $f_{pe}$ is calculated from the median $n_e$ measured by FPI-DES over the snapshot time interval. 
From May 2017 to June 2022, we identify $2.94 \times 10^5$ hmfe snapshots from all four spacecraft satisfying these criteria. For all snapshots, we high-pass filter the waveform above $f_{pe}/1.5$, so only high-frequency ${\bf E}$ fluctuations remain and contribute to the statistical results. 

\begin{figure*}[htbp!]
\begin{center}
\includegraphics[width=160mm, height=170mm]{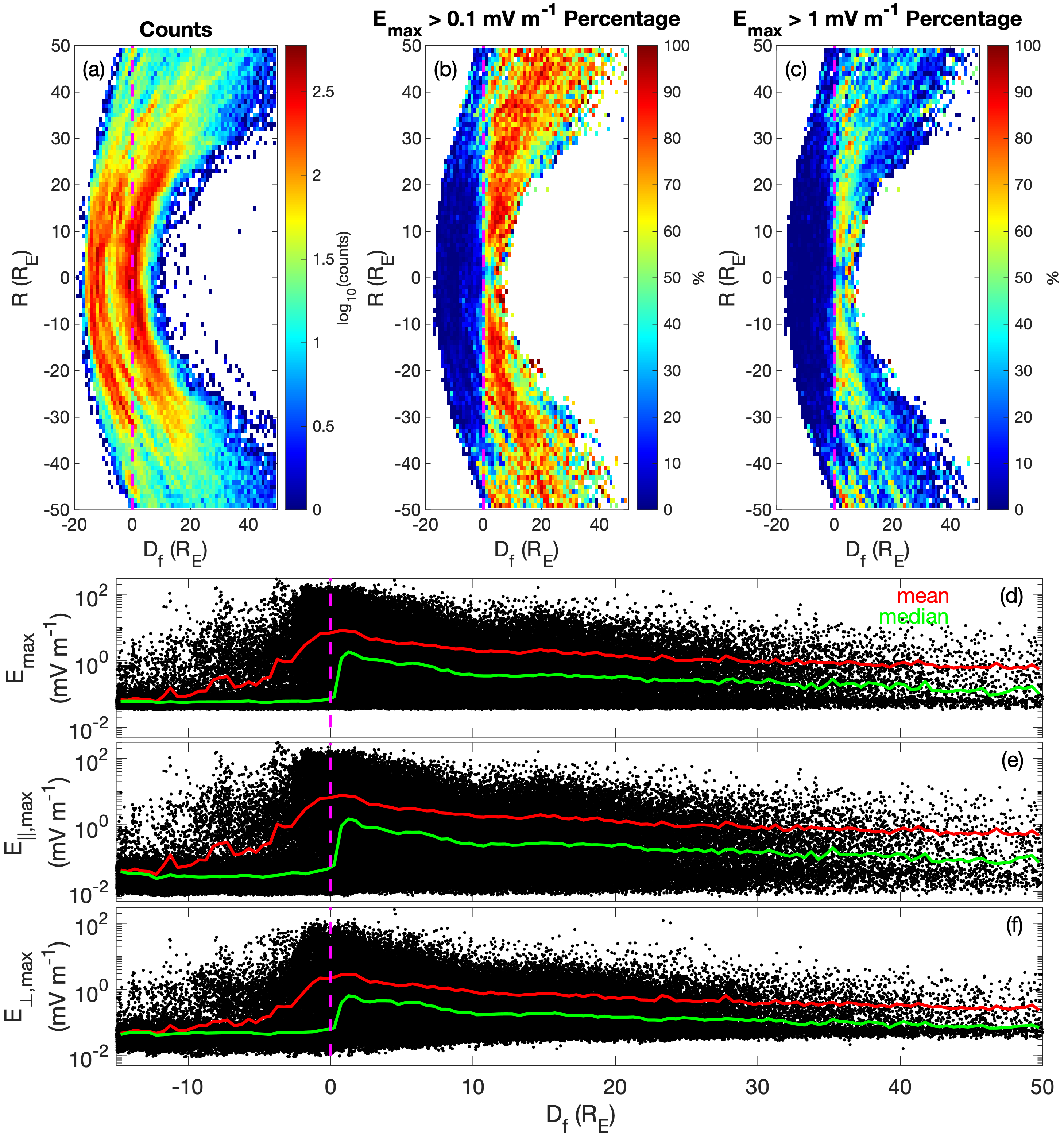}
\caption{Probability and distribution of enhanced electric fields in Earth's electron foreshock as functions of $D_f$ and $R$. (a) Two-dimensional histogram of the number of hmfe snapshots in the solar wind and electron foreshock. (b) and (c) Percentage of the snapshots satisfying $E_{\mathrm{max}} > 0.1$~mV~m$^{-1}$ and $E_{\mathrm{max}} > 1$~mV~m$^{-1}$, respectively, as functions of $D_f$ and $R$. The magenta dashed line is $D_f = 0$ indicating the boundary between the solar wind and electron foreshock. 
Scatter plots of (d) $E_{\mathrm{max}}$ versus $D_f$, (e) $E_{\parallel,\mathrm{max}}$ versus $D_f$, and (f) $E_{\perp,\mathrm{max}}$ versus $D_f$. In each panel the red and green curves are the mean and median $E_{\mathrm{max}}$ as a function of $D_f$. The magenta dashed line is $D_f = 0$. }
\label{Foreshockposstats}
\end{center}
\end{figure*}

Figure \ref{Foreshockposstats}a shows the two-dimensional histogram of the number of snapshots as a function of $D_f$ and $R$. We find that a large number of snapshots occur in both the solar wind and electron foreshock, divided by $D_f = 0$. The snapshots are observed over a wide range of $D_f$ and $R$. We note that there are two bands where the number of counts peaks. This is because burst mode intervals can only be selected when the spacecraft are operating in fast survey mode. The fast mode intervals occur during the outbound and inbound portions of the orbit around the magnetopause and bow shock, and an interval near apogee, which occurs in the solar wind when MMS's apogee is on the dayside of Earth. 

In Figure \ref{Foreshockposstats}b we plot the percentage of snapshots with maximum amplitude $E_{\mathrm{max}} > 0.1$~mV~m$^{-1}$ as a function of $D_f$ and $R$. We find enhanced ${\bf E}$ for $D_f > 0$, corresponding to regions where ${\bf B}$ is connected to the bow shock. For almost all regions where $D_f > 0$, most of the snapshots exhibit enhanced field activity. Percentages of $\gtrsim 80$~\% are common. In contrast, for $D_f < 0$ only a small fraction of snapshots satisfy $E_{\mathrm{max}} > 0.1$~mV~m$^{-1}$. Thus, when ${\bf B}$ is not connected to the bow shock, enhanced ${\bf E}$ are very rare. In Figure \ref{Foreshockposstats}c we plot the percentage of snapshots with $E_{\mathrm{max}} > 1$~mV~m$^{-1}$. In this case, we find that the probabilities are often much smaller. We find that the probabilities are typically largest for small $D_f > 0$, suggesting that the largest amplitude plasma frequency waves occur close to the boundary between the electron foreshock and the solar wind. 

To explore the dependence of wave amplitude on $D_f$, we present scatter plots of $E_{\mathrm{max}}$ for all snapshots as a function of $D_f$ in Figure \ref{Foreshockposstats}d--\ref{Foreshockposstats}f. Figure \ref{Foreshockposstats}d shows the scatterplot of $E_{\mathrm{max}}$ versus $D_f$. We find that the largest $E_{\mathrm{max}}$ are observed around $D_f = 0$. For $D_f < 0$, most snapshots have $E_{\mathrm{max}}$ at the background level, indicating no Langmuir wave activity. However, large-amplitude $E_{\mathrm{max}}$ occur for $D_f < 0$, especially when $D_f$ is close to $0$, suggesting some uncertainty in the estimation of $D_f$. For $D_f > 0$, the typical amplitude $E_{\mathrm{max}}$ tends to decrease as $D_f$ increases. We find that the largest amplitude Langmuir waves can reach $E_{\mathrm{max}} \sim 200$~mV~m$^{-1}$, which are very large and are well above the thresholds for nonlinear processes, specifically electrostatic decay. 

In Figures \ref{Foreshockposstats}e and \ref{Foreshockposstats}f we plot the scatter plots of $E_{\parallel,\mathrm{max}}$ and $E_{\perp,\mathrm{max}}$, where $E_{\parallel,\mathrm{max}}$ and $E_{\perp,\mathrm{max}}$ are the maximum $E_{\parallel}$ and $E_{\perp}$ in each snapshot. We find that the largest $E_{\parallel,\mathrm{max}}$ and $E_{\perp,\mathrm{max}}$ occur near $D_f = 0$. The distribution of $E_{\parallel,\mathrm{max}}$ is similar to $E_{\mathrm{max}}$, while $E_{\perp,\mathrm{max}}$ tends to have lower amplitudes, although the same decrease in $E_{\perp,\mathrm{max}}$ as $D_f$ increases is observed. 

In Figure \ref{Foreshockposstats}d--\ref{Foreshockposstats}f we overplot the mean and median $E_{\mathrm{max}}$, $E_{\parallel,\mathrm{max}}$, and $E_{\perp,\mathrm{max}}$. For $D_f < 0$ the medians are all at the background level, meaning most of the snapshots are characterized by negligible Langmuir wave activity. For $D_f > 0$, there is a sharp increase and peak in the medians for small $D_f$. As $D_f$ increases, there is a gradual decrease in the medians. In contrast, there is an increase in the means for $D_f < 0$ as $D_f = 0$ is approached because some of the largest amplitude Langmuir waves are observed for $D_f < 0$. This is likely due to the uncertainty in the foreshock model used to calculate $D_f$. Like the medians, the means peak around $D_f = 0$ and gradually decrease as $D_f$ increases. We conclude that the largest amplitude Langmuir waves occur around $D_f = 0$, where the fastest electron beams are expected to be found. These results are consistent with previous observations in Earth's electron foreshock \cite[]{cairns1997a,cairns2000a,sigsbee2004a}. 

\begin{figure*}[htbp!]
\begin{center}
\includegraphics[width=160mm, height=100mm]{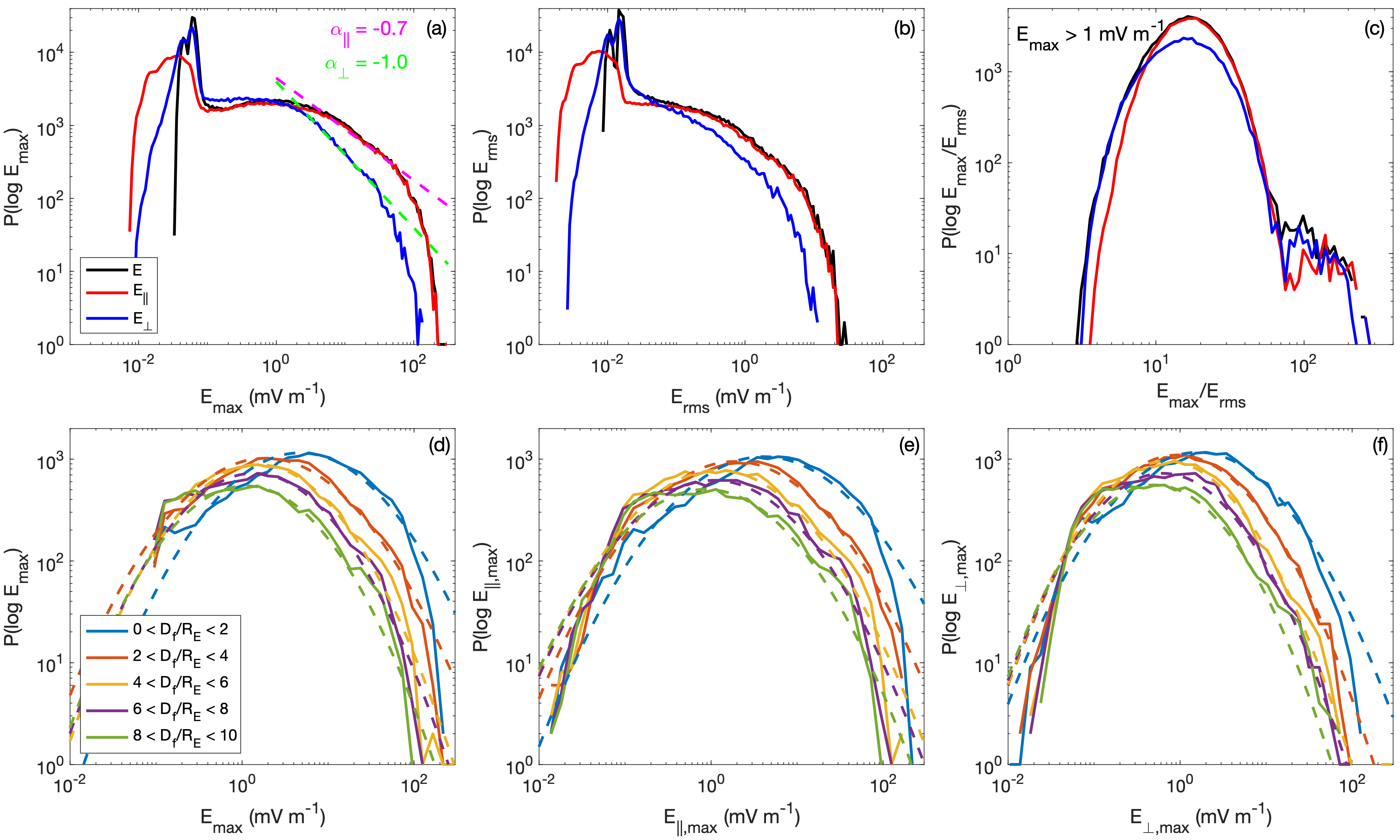}
\caption{Distributions of $\log{E}$ in counts from all solar wind snapshots and distributions as functions of $D_f$. (a) Distribution $P(\log{E})$ of $E_{\mathrm{max}}$  from each snapshot in the solar wind and foreshock. (b) Distribution $P(\log{E})$ of the root-mean-square $E_{\mathrm{rms}}$ electric field. (c) Distribution $P(\log{E_{\mathrm{max}}/E_{\mathrm{rms}}})$ versus $E_{\mathrm{max}}/E_{\mathrm{rms}}$. The black, red, and blue lines correspond to $E$, $E_{\parallel}$, and $E_{\perp}$, respectively. (d) $P(\log{E})$ of $E_{\mathrm{max}}$, (e) $P(\log{E})$ of $E_{\parallel,\mathrm{max}}$, and (f) $P(\log{E})$ of $E_{\perp,\mathrm{max}}$. The solid blue, red, gold, purple, and green solid lines are $P(\log{E})$ for $0 < D_f/R_E < 2$, $2 < D_f/R_E < 4$, $4 < D_f/R_E < 6$, $6 < D_f/R_E < 8$, $8 < D_f/R_E < 10$, respectively. The dashed lines are log-normal fits to the data using the means and standard deviations from the observed distributions. Only snapshots with $E_{\mathrm{max}} > 0.1$~mV~m$^{-1}$ are included to remove snapshots with no wave activity.}
\label{probdist1}
\end{center}
\end{figure*}

In Figure \ref{probdist1} we plot the distribution functions $P(\log{E})$ for all snapshots in the solar wind and foreshock. We plot $P(\log{E})$ for $E_{\mathrm{max}}$, $E_{\parallel,\mathrm{max}}$ and $E_{\perp,\mathrm{max}}$ in Figures \ref{probdist1}a--\ref{probdist1}c. Throughout this paper, $\log{E} = \log_{10}{E}$ is used and $E$ is in units of mV~m$^{-1}$. The peaks in $P(\log{E})$ below $E_{\mathrm{max}} = 0.1$~mV~m$^{-1}$ are due to snapshots with no enhanced wave activity and thus indicate the sensitivity of the high-frequency ${\bf E}$ measurements. For $E_{\mathrm{max}} \gtrsim 0.1$~mV~m$^{-1}$ there is a slight rise in $P(\log{E})$ with a peak at $\sim 1$~mV~m$^{-1}$. For larger $E$, there is an approximately power-law decrease in $P(\log{E})$ with $E$, and a sharp decrease in $P(\log{E})$ for $E_{\mathrm{max}} \gtrsim 100$~mV~m$^{-1}$. For $E$ and $E_{\parallel}$ the power law exponent is $\alpha_{\parallel} = -0.7$, while for $E_{\perp}$ the power law exponent is $\alpha_{\perp} = -1$. The power law decrease in $P(\log{E})$ is consistent with previous observations in the foreshock \cite[]{bale1997,cairns1997b}. In Figure \ref{probdist1}b we plot $P(\log{E})$ for the root-mean-square electric field $E_{\mathrm{rms}}$ of all snapshots in the solar wind and foreshock. We find that $P(\log{E})$ of $E_{\mathrm{rms}}$, $E_{\parallel,\mathrm{rms}}$ and $E_{\perp,\mathrm{rms}}$ exhibit qualitatively similar distributions to Figure \ref{probdist1}a, although the power-law dependence is less pronounced, and $P(\log{E})$ is shifted to lower $E$. 

In Figure \ref{probdist1}c we plot the probability distribution  $P(\log{E_{\mathrm{max}}/E_{\mathrm{rms}}})$ versus $E_{\mathrm{max}}/E_{\mathrm{rms}}$ for each of the snapshots with $E_{\mathrm{max}} > 1$~mV~m$^{-1}$. We find similar distributions for $E$, $E_{\parallel}$, and $E_{\perp}$, with peaks in $P(\log{E_{\mathrm{max}}/E_{\mathrm{rms}}})$ at $E_{\mathrm{max}}/E_{\mathrm{rms}} \approx 16$. This distribution shows that the Langmuir/Z-mode waves are generally bursty, with localized enhancements in $E$. This bursty behavior is observed for the examples in Figure \ref{LZexamples} and applies generally to Langmuir/Z-mode waves in the electron foreshock. The values $E_{\mathrm{max}}/E_{\mathrm{rms}}$ are substantially larger than those found by \cite{bale1997} in the electron foreshock. However, we note that \cite{bale1997} relied on substantially shorter snapshots and different triggers for the waveforms from the Wind spacecraft, compared with those used here. 

Finally, we consider $P(\log{E})$ for different ranges of $D_f$ since the typical amplitude of waves varies with $D_f$ (Figures \ref{probdist1}d--\ref{probdist1}f). Figures \ref{probdist1}d--\ref{probdist1}f show $P(\log{E})$ at different ranges of $D_f$ for $E_{\mathrm{max}}$, $E_{\parallel,\mathrm{max}}$, and $E_{\perp,\mathrm{max}}$.  We only consider snapshots with $E_{\mathrm{max}} > 0.1$~mV~m$^{-1}$ to exclude snapshots where no waves are observed. In contrast to the distribution in Figure \ref{probdist1}a over all $D_f$, we do not observe any clear power law scaling of $P(\log{E})$. For $E_{\mathrm{max}}$, $E_{\parallel,\mathrm{max}}$, and $E_{\perp,\mathrm{max}}$, we find that the distributions peak at lower $E_{\mathrm{max}}$ as $D_f$ increases. This corresponds to the statistical decrease in amplitude of the waves as $D_f$ increases, as seen in Figure \ref{Foreshockposstats}. 

When a narrow range of $D_f$ is considered, $P(\log{E})$ is similar to a normal distribution as a function of $\log{E}$. The dashed curves in Figures \ref{probdist1}d--\ref{probdist1}f show normal distributions calculated from the means and standard deviations of the observed distributions. A normal distribution in $\log{E}$ is predicted by Stochastic Growth Theory (SGT) \cite[]{robinson1992a} and observed in previous studies of the electron foreshock \cite[]{cairns1997b,cairns1999,sigsbee2004a}. 

In summary, large-amplitude Langmuir and beam-mode waves develop in the electron foreshock, with the largest amplitude waves developing around $D_f = 0$ where the fastest electron beams are expected. Nevertheless, enhanced electric fields can persist to large distances downstream of $D_f = 0$. Although the largest component of the electric field is aligned with ${\bf B}$, large-amplitude electric fields perpendicular to ${\bf B}$ are frequently observed.

\section{Field statistics of Langmuir waveforms} \label{fieldstatssec}
We now investigate the statistical properties of the Langmuir waveforms using probability distribution functions of the envelope electric field from individual hmfe snapshots. 

\subsection{Stochastic Growth Theory and probability distributions}
We compare the observed electric field distributions with the predictions of both linear and nonlinear Stochastic Growth Theory (SGT). In SGT, the plasma is assumed to be close to marginal stability and inhomogeneous. The plasma inhomogeneities result in localized regions of damping and growth. The magnitude of the electric field is given by $E(t) = E_0 \mathrm{e}^{G(t)}$, where $E(t)$ is the observed electric field, $E_0$ is a reference electric field, and $G$ is the gain or number of e-foldings. The gain is given by
\begin{equation}
G(t) = \int_{- \infty}^{t} dt' \gamma(t') \approx \sum_i \gamma(t_i) \Delta t_i,
\label{gaineq}
\end{equation}
where $\gamma$ is the growth or damping rate. By assuming that $\gamma$ is a random variable, $G$ will undergo a random walk. Assuming a large number of fluctuations in $\gamma$ are encountered, the distribution of $G$ will approach a normal distribution according to the Central Limit Theorem. Thus, the probability distribution $P(\log{E})$ is predicted to be a normal distribution as a function of $\log{E}$, and is given by \cite[]{robinson1992a}
\begin{equation}
P(\log{E}) = \frac{1}{\sigma \sqrt{2 \pi}} \exp{\left(- \frac{(\log{E} - \mu)^2}{2 \sigma^2} \right)},
\label{sgteqlin}
\end{equation}
where $\mu$ is the mean and $\sigma$ is the standard deviation of the distribution. In Equation (\ref{sgteqlin}), we have assumed that the waves are purely linear. The effect of nonlinear processes will be to modify $P(\log{E})$ at large $\log{E}$.  In particular, for electrostatic decay, a rapid drop in $P(\log{E})$ for $E$ above the threshold of electrostatic decay is predicted due to Langmuir waves being shifted out of resonance with the electron beam, arresting their growth. To model the effect of electrostatic decay, a cutoff $E_c$ is introduced and modeled as an absorbing boundary. The distribution function is given by \cite[]{robinson1993a,robinson1995,cairns2001}
\begin{multline}
P(\log{E}) = \frac{1}{\sigma_{*} A \sqrt{2 \pi}} \Bigg[ \exp{\left( \frac{-(\log{E} - \mu_{*})^2}{2 \sigma_{*}^2} \right)}  \\ - \exp{\left( \frac{-(2 \log{E_c} - \log{E} - \mu_{*})^2}{2 \sigma_{*}^2} \right)} \Bigg],
\label{sgteqnl}
\end{multline}
where $A = \mathrm{erf}[(\log{E_c} - \mu_{*})/(\sigma_{*} \sqrt{2})]$ is the normalization for the distribution and $\mathrm{erf}$ is the error function. For $E > E_c$, $P(\log{E}) = 0$ is used, and equation (\ref{sgteqnl}) requires $\log{E_c} > \mu_{*}$. Here we introduce $\mu_{*}$ and $\sigma_{*}$ because they differ from $\mu$ and $\sigma$ calculated from the moments of the distribution.

We calculate the probability distributions of $\log E$ from the large-amplitude Langmuir wave snapshots in the electron foreshock (section \ref{wavepropssec}). Since SGT applies to the amplitude of the waves,  we use the envelope electric field $E_{\mathrm{env}}$ of ${\bf E}$ rather than the fluctuating waveform itself. To compute $E_{\mathrm{env}}$, we apply a Hilbert transform to ${\bf E}$ to produce ${\bf E}_H$, which has a phase shift of $90^{\circ}$ from the original signals. Each component of ${\bf E}_{\mathrm{env}}$ is calculated using $E_{\mathrm{env}} = \sqrt{E^2 + E_H^2}$, which is then smoothed to remove any residual fluctuations close to the wave frequency. 

For the waveforms in section \ref{wavepropssec}, we calculate the mean frequency and spectral width of the waves in each snapshot from the power spectrum using
\begin{equation}
f_0 = \frac{ \sum \, f E(f)^2 \Delta f}{\sum E(f)^2 \Delta f},
\label{f0eq}
\end{equation}
\begin{equation}
\delta f^2 = \frac{ \sum \, (f- f_0)^2 E(f)^2 \Delta f}{\sum E(f)^2 \Delta f},
\label{deltafeq}
\end{equation}
where $f_0$ and $\delta f$ are the mean frequency and spectral width of the waves and $\Delta f$ is the frequency resolution. We apply equations (\ref{f0eq}) and (\ref{deltafeq}) to the Fourier transforms of the entire snapshots. To primarily focus on Langmuir/Z-mode waves, rather than more broadband beam-mode waves, we select snapshots satisfying $\delta f/f_0 < 0.1$. Additionally, we only consider waveforms with $min(E_{\mathrm{env}}) > 0.02$~mV~m$^{-1}$ and median $E_{\mathrm{env}}$ exceeding $0.2$~mV~m$^{-1}$. These conditions are chosen to ensure that snapshots with extended periods of time with little or no Langmuir wave activity are not included in the statistics. When there is no Langmuir wave activity, the values of $E_{env}$ simply correspond to instrumental background, which leads to very skewed $P(\log E)$. After applying these constraints, we are left with 5,129 Langmuir wave snapshots from the total 43,664 in the electron foreshock with $E_{\mathrm{max}} > 5$~mV~m$^{-1}$ (section \ref{wavepropssec}). To compute $P(\log E)$, we calculate the histogram of $\log{E}$ over the domain $\log{E_{\mathrm{min}}}$--$\log{E_{\mathrm{max}}}$ using 30 evenly spaced bins, where $E_{\mathrm{min}}$ and $E_{\mathrm{max}}$ are the minimum and maximum electric field strengths over the snapshot. The histogram is then normalized so $\int P(\log{E}) d \log{E} = 1$. 

We fit equations (\ref{sgteqlin}) and (\ref{sgteqnl}) to the observed $P(\log{E})$ to determine which equation provides the best fit to the data, and hence, if there is evidence of electrostatic decay occurring. To find the best fit of the model $P(\log{E})$ to the observations, we use a Nelder-Mead method to minimize 
\begin{equation}
\chi^2 = \sum \frac{\left[ P_{obs}(\log E) - P_{model}(\log E) \right]^2}{P_{model}(\log E)},
\label{chi2eq}
\end{equation}
where $P_{obs}$ and $P_{model}$ refer to the observed $P(\log{E})$ and modeled probability distributions [equations (\ref{sgteqlin}) and (\ref{sgteqnl})]. We then calculate the reduced $\chi^2$, $\chi_r^2 = \chi^2/\nu$, where $\nu = N - k -1$ is the number of degrees of freedom, $N = 30$ is the number of bins, and $k$ is the number of parameters in the fitted equations. Since the nonlinear SGT case has one additional free parameter compared with linear SGT, for equal $\chi^2$, $\chi_r^2$ will be smaller for linear SGT.

In addition to computing $P(\log{E})$, we also calculate the moments of the observed distribution of $\log{E}$ to quantify the distributions. The moments are given by
\begin{equation}
\mu_1 = \frac{1}{n} \sum_{i = 1}^n \log{E_i}, 
\label{mu0eq}
\end{equation}
\begin{equation}
\mu_j = \frac{1}{n} \sum_{i = 1}^n (\log{E_i} - \mu_1)^j, 
\label{mujeq}
\end{equation}
where $j > 1$ is the order of the moment, and $n$ is the number of points of $E_{\mathrm{env}}$ in each snapshot. Here, $\mu = \mu_1$ is the mean of the distribution, and $\sigma = \sqrt{\mu_2}$ is the standard deviation. For MMS's hmfe snapshots, $n$ is nominally $131,072$.
%For nominal solar wind conditions, the $2$~s snapshot durations correspond to spatial scales of $l \approx 800$~km $\approx 6 \times 10^4 \, \lambda_D$, where $\lambda_D$ is the Debye length.   

From these moments, we calculate the square of the skewness $\beta_1$ and kurtosis $\beta_2$ of the distributions, which are given by 
\begin{equation}
\beta_1 = \frac{\mu_3^2}{\mu_2^3},
\beta_2 = \frac{\mu_4}{\mu_2^2}.
\label{beta12}
\end{equation}
The skewness squared $\beta_1$ provides a measure of the asymmetry of $P(\log{E})$ about the mean $\mu$, while the kurtosis $\beta_2$ is a measure of the tailedness of the distribution far from the mean of the distribution. For a normal distribution, and hence the linear SGT prediction [equation (\ref{sgteqlin})], $(\beta_1,\beta_2) = (0,3)$ is predicted. For the case of homogeneous linear growth and damping, a uniform distribution is expected \cite[]{cairns2001}. Continuous uniform distributions are characterized by $(\beta_1,\beta_2) = (0,1.8)$. Values of $\beta_1$ and $\beta_2$ that differ from the $(\beta_1,\beta_2) = (0,3)$ prediction may suggest that linear SGT does not apply to the observed waveform, or that nonlinear processes could be active.

\begin{figure}[htbp!]
\begin{center}
\includegraphics[width=90mm, height=40mm]{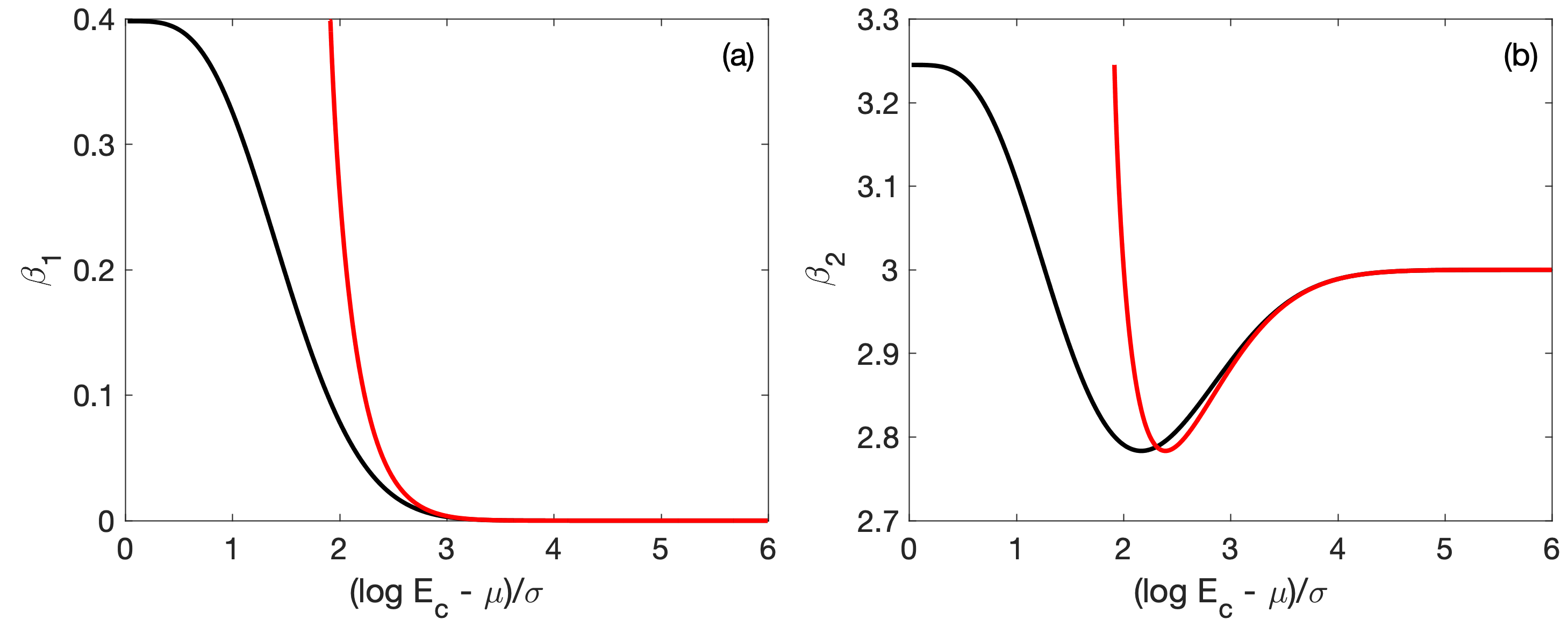}
\caption{Plots of (a) $\beta_1$ versus $(\log{E_c} - \mu)/\sigma$ and (b) 
$\beta_2$ versus $(\log{E_c} - \mu)/\sigma$. The black curves are for $(\log{E_c} - \mu_{*})/\sigma_{*}$ where 
$\mu_{*}$ and $\sigma_{*}$ are input parameters in equation (\ref{sgteqnl}) and the red curves are for $(\log{E_c} - \mu)/\sigma$ using $\mu$ and $\sigma$ calculated from the moments of the distribution function.}
\label{sgtnl}
\end{center}
\end{figure}

To illustrate how $\beta_1$ and $\beta_2$ vary due to electrostatic decay, we calculate $\beta_1$ and $\beta_2$ from equation (\ref{sgteqnl}) where $\log E_c$ is a free parameter. In Figure \ref{sgtnl}, we plot $\beta_1$ and $\beta_2$ as functions of $(\log{E_c} - \mu)/\sigma$. The black curve corresponds to $(\log{E_c} - \mu_{*})/\sigma_{*}$ where $\mu_{*}$ and $\sigma_{*}$ are parameters in equation (\ref{sgteqnl}) and the red curve corresponds to $(\log{E_c} - \mu)/\sigma$ using $\mu$ and $\sigma$ calculated from the moments of the distribution function. Figure \ref{sgtnl}a shows that $\beta_1$ has a maximum value of $0.4$ as $\log{E_c}$ approaches $\mu_{*}$. As $\log{E_c}$ becomes much larger than $\mu$, $\beta_1$ approaches $0$. Figure \ref{sgtnl}b shows that $\beta_2$ can range from $\approx 2.8$ to $\approx 3.25$. In particular, for the lowest $\log{E_c}$, $\beta_2 > 3$, while for $\log{E_c} > \mu + 2 \sigma$, $\beta_2 < 3$ and approaches $3$ as $\log{E_c}$ becomes large. 

%From the observed $P(\log{E})$ where nonlinear SGT provides a better fit to the data we find that $(\log{E_c} - \mu)/\sigma$ is typically between $2$ and $3$ with a median of $2.3$. For this range $0 \lesssim \beta_1 \lesssim 0.4$ is expected. From observations we obtain a median $\beta_1 = 0.06$, which is also the value found for all snapshots. However, we find that $\beta_1$ tends to decrease as $(\log{E_c} - \mu)/\sigma$ consistent with Figure \ref{sgtnl}a. For $\beta_2$ values less than $3$ are generally expected for these distributions. However, we find that for the distributions where nonlinear SGT provides a better fit to the data the median $\beta_2$ is $2.5$ compared with the median $\beta_2 = 2.8$ of all snapshots analyzed. Overall, we conclude that nonlinear effects can in part explain some of the statistical results. However, other effects are needed to fully explain the statistical results. 

Previous studies of the statistical distributions of Langmuir waves have argued that the Pearson system of distribution functions \cite[]{pearson1895} better models the distribution of Langmuir waves \cite[]{krasnoselskikh2007,musatenko2007,vidojevic2014,voshchepynets2017}. The Pearson system of distribution functions is defined by 
\begin{equation}
\frac{d f(x)}{d x} = -\frac{x - a}{d x^2 + c x + b} f(x), 
\label{pearsoneq}
\end{equation}
where $a$, $b$, $c$, $d$ are coefficients. These coefficients can be calculated from $\mu$, $\sigma$, $\beta_1$, and $\beta_2$, and determine the type of the distribution function \cite[]{podladchikova2003}. In the following subsections, we will compare the calculated $\beta_1$ and $\beta_2$ to the predictions of linear and nonlinear SGT and the system of Pearson distributions. 

\begin{figure*}[htbp!]
\begin{center}
\includegraphics[width=160mm, height=100mm]{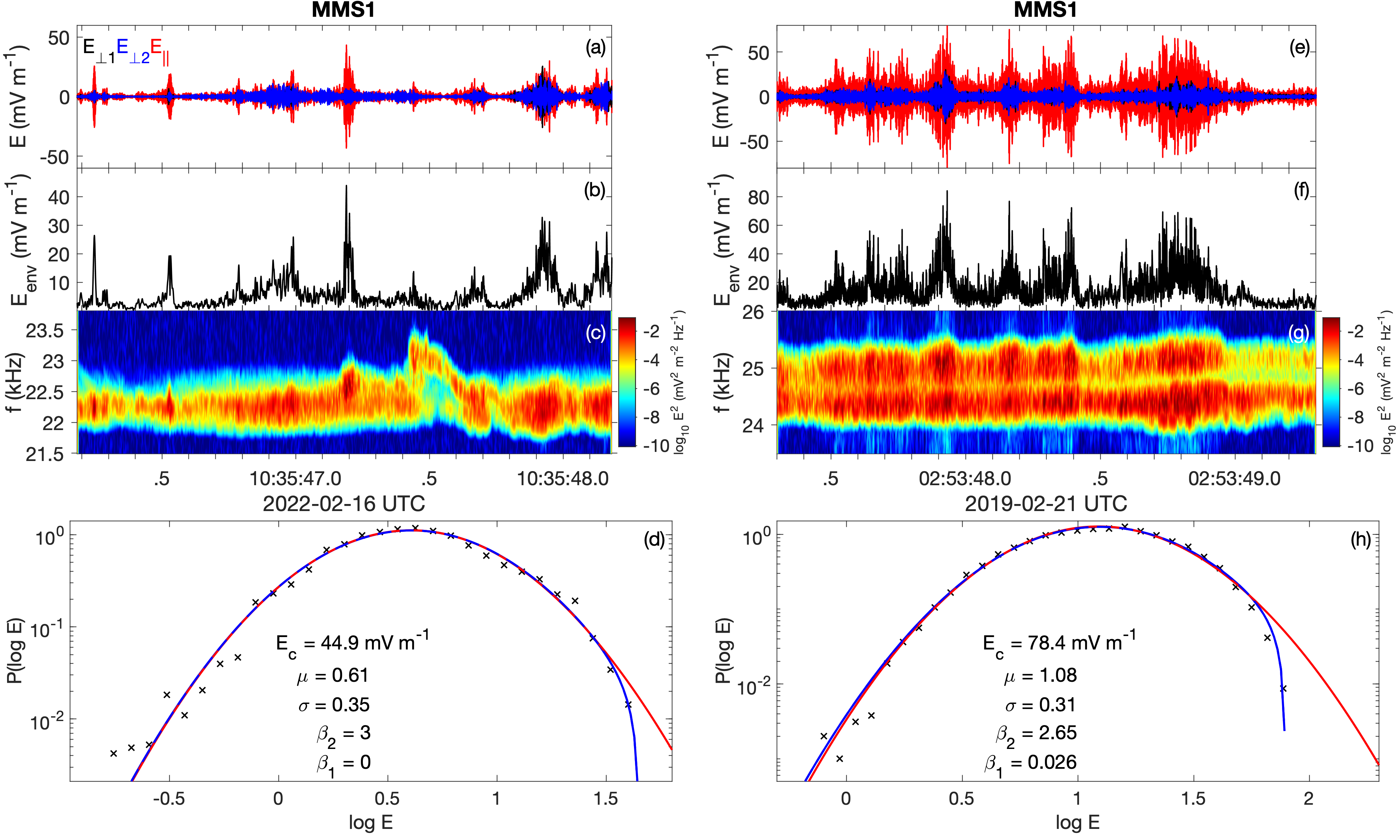}
\caption{Two examples of Langmuir/Z-mode waves and their associated probability distributions $P(\log{E})$.  (a)--(d) Langmuir waveform observed by MMS1 on 2018 February 18 and (e)--(f) Langmuir waveform observed by MMS1 on 2019 February 21. (a) and (e) ${\bf E}$ in field-aligned coordinates. (b) and (f) $E_{\mathrm{env}}$. (c) and (g) Frequency-time power spectra of the Langmuir waves. (d) and (h) $P(\log{E})$ calculated from $E_{\mathrm{env}}$. The properties of $P(\log{E})$ are given in the panel. The red and blue lines show the best fits of equations (\ref{sgteqlin}) and (\ref{sgteqnl}) to the data. }
\label{sgtexamples}
\end{center}
\end{figure*}

\subsection{Probability distributions of the total electric field} \label{sgtEtot}
We first consider $P(\log{E})$ of the total electric field envelope $E_{env}$. Figure \ref{sgtexamples} shows two hmfe snapshots of Langmuir waveforms and the associated $P(\log{E})$. Figures \ref{sgtexamples}a--\ref{sgtexamples}d show an example of bursty Langmuir/Z-mode waves. The waveform and $E_{\mathrm{env}}$ exhibit rapid short-scale variations in amplitude, and large $E_{\perp}$ are observed throughout the interval, suggesting Z-mode waves are present. The bursts in ${\bf E}$ have time scales of a few tens of milliseconds, with variable spacing between them. This is also seen in the spectrogram of ${\bf E}$ (Figure \ref{sgtexamples}c), where there are rapid bursts of power over a narrow frequency range. Throughout the interval, there is some variation in the frequency of the Langmuir waves, although there is little evidence of the two distinct simultaneously spectral peaks expected from electrostatic decay. 

In Figure \ref{sgtexamples}d, we plot $P(\log{E})$ calculated from $E_{\mathrm{env}}$, along with the properties of the distribution. The best fits of equations (\ref{sgteqlin}) and (\ref{sgteqnl}) to the data are overplotted. We find good agreement between the observed $P(\log{E})$ and the prediction from linear SGT. The fit of Equation (\ref{sgteqnl}) deviates from the linear SGT prediction for large $\log{E}$. From equation (\ref{chi2eq}) we obtain $\chi_r^2 = 7.6 \times 10^{-3}$ and $7.7 \times 10^{-3}$ for the linear and nonlinear SGT fits. Thus, we find that $\chi_r^2$ is smaller for linear SGT, meaning that nonlinear SGT does not provide a significantly improved fit. This suggests that nonlinear processes are not occurring or are not significant enough to contribute to the $P(\log{E})$. We calculate $(\beta_1, \beta_2) = (0.00,3.00)$, which matches the prediction of linear SGT and is inconsistent with homogeneous growth or damping. This example provides evidence that the Langmuir/Z-mode waves undergo stochastic growth and damping.

Figures \ref{sgtexamples}e--\ref{sgtexamples}f show an intense Langmuir wave snapshot, characterized by two distinct spectral peaks (Figure \ref{sgtexamples}g), which persist throughout the hmfe snapshot. On average, the two peaks are separated in frequency by $\approx 750$~Hz. In this example, ${\bf E}$ is primarily aligned with ${\bf B}$. Like the previous event, the amplitude of the waves varies rapidly in time (Figures \ref{sgtexamples}e and \ref{sgtexamples}f). The most rapid variations are due to the beating between the two waves with distinct frequencies. However, Figure \ref{sgtexamples}g shows that the power associated with each spectral peak is also highly variable in time. In Figure \ref{sgtexamples}h, we plot $P(\log{E})$ calculated from $E_{\mathrm{env}}$. Overall, $P(\log{E})$ remains close to the linear SGT prediction, except at the highest $\log{E}$ where nonlinear SGT provides a better fit to the data. We calculate $\chi_r^2 = 5.4 \times 10^{-3}$ and $2.9 \times 10^{-3}$ for the linear and nonlinear SGT fits. Based on the nonlinear SGT fit, $E_c = 78$~mV~m$^{-1}$, which is comparable to $E_{\mathrm{max}}$. For this snapshot, we calculate $(\beta_1, \beta_2) = (0.03,2.64)$, which indicates a deviation from a normal distribution predicted by linear SGT, and $\beta_2$ is smaller than predicted by equation (\ref{sgteqnl}). The slight skewness and decreased kurtosis are in part due to the sharp drop in $P(\log{E})$ at the highest $\log{E}$. The presence of two distinct spectral peaks, and $P(\log{E})$ having better agreement with nonlinear SGT, is consistent with electrostatic decay occurring in this event. 

\begin{figure}[htbp!]
\begin{center}
\includegraphics[width=90mm, height=80mm]{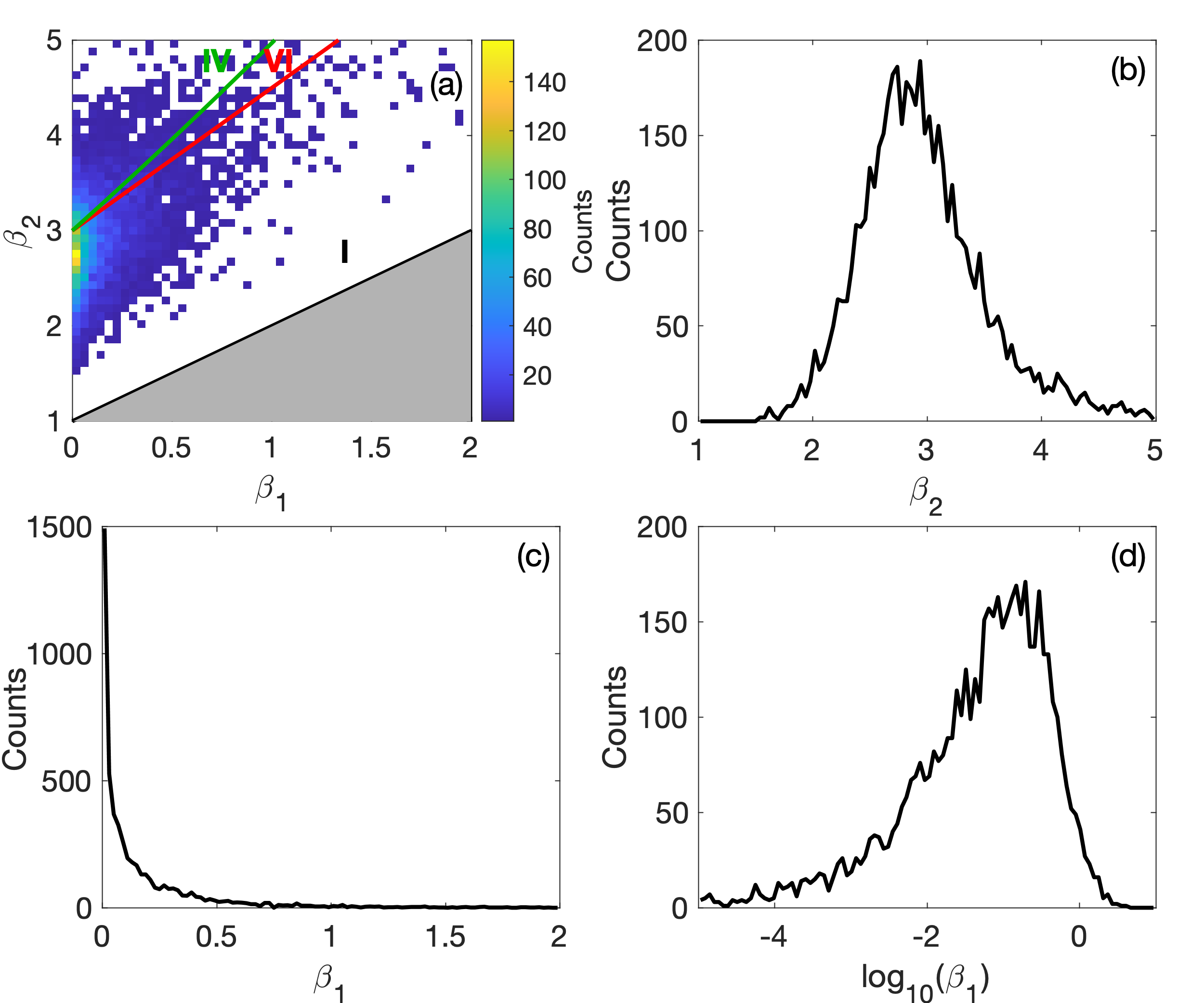}
\caption{Statistics of $\beta_1$ and $\beta_2$ calculated from $E_{\mathrm{env}}$. (a) Two-dimensional histogram of $\beta_1$ and $\beta_2$. The gray-shaded region corresponds to $\beta_2 < \beta_1+1$, which is not possible. The red and green lines mark indicate the boundaries between type I and type VI, and type VI and type IV Pearson distributions, respectively. (b) Histogram of $\beta_2$. 
(c) and (d) Histograms of $\beta_1$ and $\log_{10} \beta_1$. }
\label{beta12stats}
\end{center}
\end{figure}

We now investigate statistically $P(\log{E})$ of the total electric field envelope. For each of the 5,129 selected snapshots, we calculate $P(\log{E})$ along with $\beta_1$ and $\beta_2$. We compare the $\beta_1$ and $\beta_2$ with SGT and discuss the type of Pearson distribution expected. The type of Pearson distribution is determined by \cite[]{nagahara2004}
\begin{equation}
\kappa = \frac{\beta_1 (\beta_2 + 3)^2}{4 (2 \beta_2 - 3 \beta_1 - 6) (4 \beta_2 - 3 \beta_1)},
\label{kappaeq}
\end{equation}
where $\kappa < 0$ corresponds to type I (beta distribution), $0 < \kappa < 1$ corresponds to type IV (not a standard distribution function), and  $\kappa > 1$ corresponds to type VI (beta prime distribution). 

Figure \ref{beta12stats} shows the statistical results of $\beta_1$ and $\beta_2$ for the analyzed snapshots. In Figure \ref{beta12stats}a we plot the two-dimensional histogram of $\beta_1$ and $\beta_2$ for all snapshots. The gray-shaded region corresponds to $\beta_2 < \beta_1 + 1$, which is impossible to reach, and the lines mark the boundaries between different function types in the Pearson system. Most distributions are characterized by $\beta_1 \ll 1$ and $\beta_2$ close to but smaller than $3$. There are negligible events with $(\beta_1, \beta_2) \approx (0.0,1.8)$, indicating that uniform distributions are unlikely to be observed. We find that 75~\% of the waveforms are type I, 22~\% are type IV, and 4~\% are type VI. 

Figure \ref{beta12stats}b shows the histogram of $\beta_2$. We find that most of the snapshots have $\beta_2 \lesssim 3$. The distribution has a mean and median of $3.0$ and $2.9$, indicating good agreement with the predictions of linear and nonlinear SGT. Figures \ref{beta12stats}c and \ref{beta12stats}d show the distributions of $\beta_1$ and $\log{\beta_1}$. We find that $\beta_1$ is small with a mean and median of $0.18$ and $0.07$. These values of $\beta_1$ are consistent with the predictions of nonlinear SGT, with $\beta_1$ potentially increasing from $0$ due to electrostatic decay. Overall, the results suggest that the distributions of $\log{E}$ tend to be relatively symmetric about their mean $\log{E}$. 

For all the snapshots in Figure \ref{beta12stats} we fit equations (\ref{sgteqlin}) and (\ref{sgteqnl}) to each of the field distributions. We find that $33 \, \%$ of the snapshots are better fitted by the nonlinear SGT prediction. Specifically, the nonlinear model has a lower $\chi_r^2$ than the linear model (see Appendix \ref{app1} for details). In the cases where nonlinear SGT better models the data, we find that the median $E_c$ is $31$~mV~m$^{-1}$. The 10th and 90th percentiles of $E_c$ are $8$~mV~m$^{-1}$ and $92$~mV~m$^{-1}$, respectively, which suggests that the threshold for nonlinear decay is typically tens of mV~m$^{-1}$ in the electron foreshock. We find that the Pearson system of distributions provide better fits to the observed $P(\log E)$ in $95 \, \%$ of cases compared with linear and nonlinear SGT. This is primarily due to equation (\ref{pearsoneq}) describing a wide variety of distributions, including the SGT prediction (see Appendix \ref{app1} for details). 

For SGT including electrostatic decay, we predict $\beta_1 < 0.4$ and $2.78 < \beta_2 < 3.25$. We find that $33\,\%$ of the waveforms satisfy these criteria. We note that errors in $\beta_1$ and $\beta_2$ can be calculated from the higher-order moments of the distribution \cite[]{pearson1902}. We find that the typical errors in $\beta_1$ and $\beta_2$ are $\lesssim 0.01$, which when included do not significantly change the statistical results. This suggests that linear and nonlinear SGT often explain the observed $P(\log{E})$. 

The fact that $33$~\% of the waveforms have field statistics more consistent with nonlinear SGT, along with the observations of two distinct spectral peaks in some snapshots, suggests that three-wave decay is common in the electron foreshock. The presence of electrostatic decay may, in part, account for the deviation of $\beta_2$ from $3$ and for why $\beta_1 > 0$. 

\subsection{Probability distributions of the parallel and perpendicular electric field} \label{sgtEparperp}
\begin{figure}[htbp!]
\begin{center}
\includegraphics[width=90mm, height=110mm]{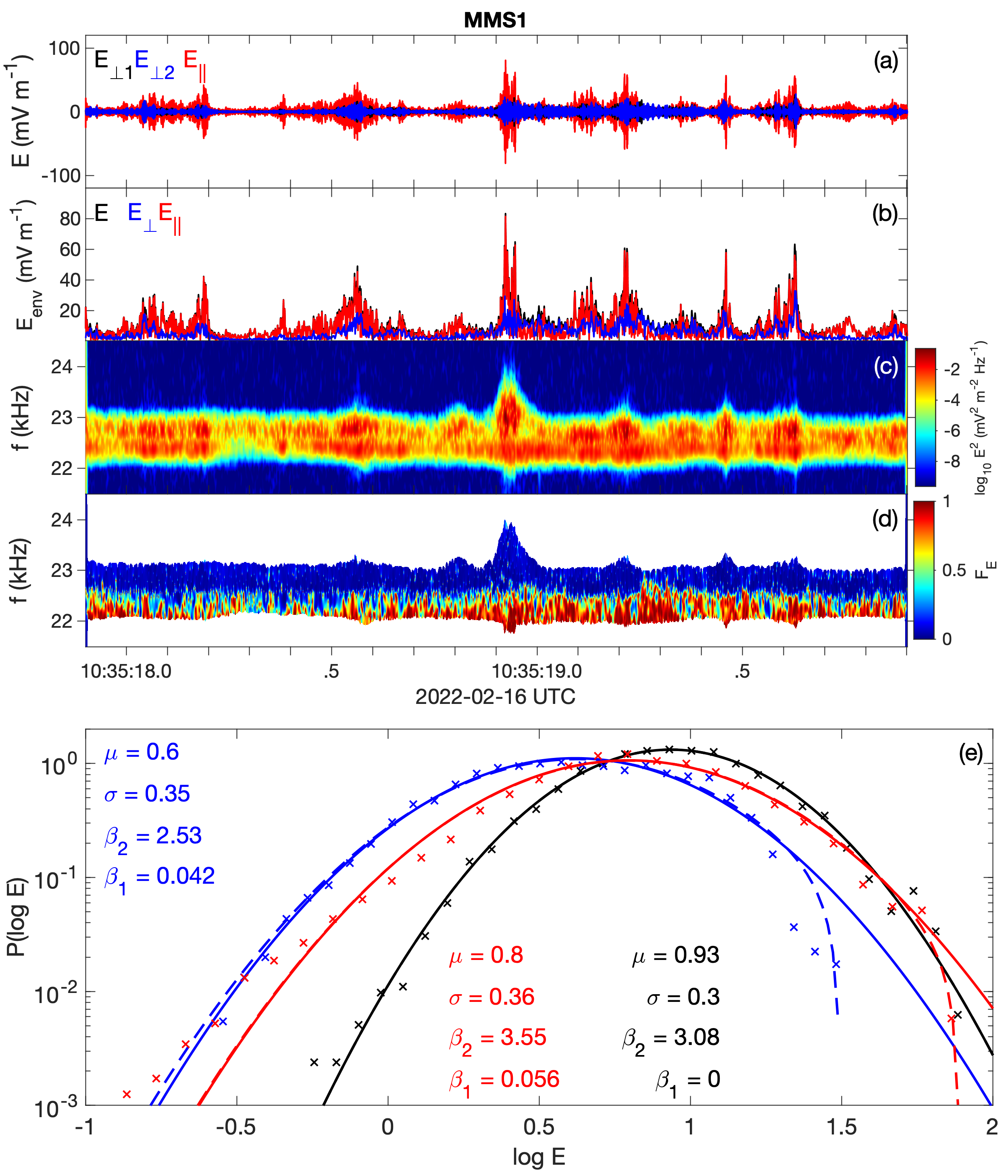}
\caption{Waveform and field statistics of a Langmuir/Z-mode snapshot with large perpendicular electric fields. (a) ${\bf E}$ in field-aligned coordinates. (b)  Envelope fields of $E$ (black), $E_{\parallel}$ (red), $E_{\perp}$ (red). (c) Frequency-time spectrogram of ${\bf E}$. (d) Spectrogram of $F_E$. (e) $P(\log{E})$ of $E_{\mathrm{env}}$ (black crosses), $E_{\parallel,\mathrm{env}}$ (red crosses), and $E_{\perp,\mathrm{env}}$ (blue crosses). The solid lines are the best fits of equation (\ref{sgteqlin}) to $P(\log{E})$ and the dashed lines are the best fits of equation (\ref{sgteqnl}) to $P(\log{E})$. The properties of $P(\log{E})$ are printed in panel (e) for $E_{\mathrm{env}}$, $E_{\parallel,\mathrm{env}}$, and $E_{\perp,\mathrm{env}}$. }
\label{egsgtparperp}
\end{center}
\end{figure}

We now consider the statistical distributions of $E_{\parallel}$ and $E_{\perp}$. We apply the criteria $min(E_{\parallel,\mathrm{env}}), min(E_{\perp,\mathrm{env}}) > 0.01$~mV~m$^{-1}$ and median $E_{\parallel,\mathrm{env}}$ and $E_{\perp,\mathrm{env}}$ exceeding $0.1$~mV~m$^{-1}$, where $E_{\parallel,\mathrm{env}}$ and $E_{\perp,\mathrm{env}}$ are the envelopes of $E_{\parallel}$ and $E_{\perp}$. After applying these criteria, we obtain 2,544 snapshots from the 5,129 snapshots in section \ref{sgtEtot}. As an example, Figure \ref{egsgtparperp} shows $E_{\mathrm{env}}$ of $E$, $E_{\parallel}$, and $E_{\perp}$, and the associated $P(\log{E})$. The waveform of ${\bf E}$ (Figure \ref{egsgtparperp}a) and $E_{\mathrm{env}}$ (Figure \ref{egsgtparperp}b) are characterized by rapid localized fluctuations in amplitude. We find that large $E_{\perp}$ is observed throughout the snapshot, with a peak of $30$~mV~m$^{-1}$, although $E_{\parallel}$ is typically the dominant component of ${\bf E}$ with $F_E = 0.25$. This can be seen in Figure \ref{egsgtparperp}b, where $E_{\mathrm{env}} \approx E_{\parallel,\mathrm{env}}$. Figure \ref{egsgtparperp}c shows that the waves are characterized by two distinct spectral peaks, which extend across the snapshot interval. The lower spectral peak remains at a relatively constant frequency, while there are some variations in frequency for the upper spectral peak. This is most evident in the center of the snapshot, where there is an increase in frequency, where ${\bf E}$ peaks. Figure \ref{egsgtparperp}d shows the spectrogram of $F_E$. The upper spectral peak is characterized by $F_E \approx 0$, corresponding to field-aligned Langmuir waves, while the lower spectral peak often has $F_E \approx 1$, indicating Z-mode waves. Overall, the wave properties are similar to those in Figure \ref{LZexamples}j--\ref{LZexamples}l. 

In Figure \ref{egsgtparperp}e we plot $P(\log{E})$ for $E_{\mathrm{env}}$, $E_{\parallel,\mathrm{env}}$, $E_{\perp,\mathrm{env}}$. For $E_{\mathrm{env}}$ we find that $P(\log{E})$ is well modeled by linear SGT [equation (\ref{sgteqlin})] with $\beta_1 = 0$ and $\beta_2 = 3.08$, despite two distinct spectral peaks being present, suggestive of three-wave decay. Nonlinear SGT does not provide a better fit, with larger $\chi_r^2$. For $E_{\parallel,\mathrm{env}}$ and $E_{\perp,\mathrm{env}}$ we find larger deviations from linear SGT. Specifically, for $E_{\parallel,\mathrm{env}}$ there is a tail in $P(\log{E})$ for small $\log{E}$, while for $E_{\perp,\mathrm{env}}$ there is a sharp drop in $P(\log{E})$ for large $\log{E}$. For $E_{\parallel,env}$, $\chi_r^2$ is reduced for linear SGT, so nonlinear SGT does not significantly improve the fit. For $E_{\perp,env}$, $\chi_r^2$ is reduced for nonlinear SGT; however, it is not clear if the SGT predictions apply to $E_{\perp}$ or Z-mode waves.

\begin{figure}[htbp!]
\begin{center}
\includegraphics[width=90mm, height=75mm]{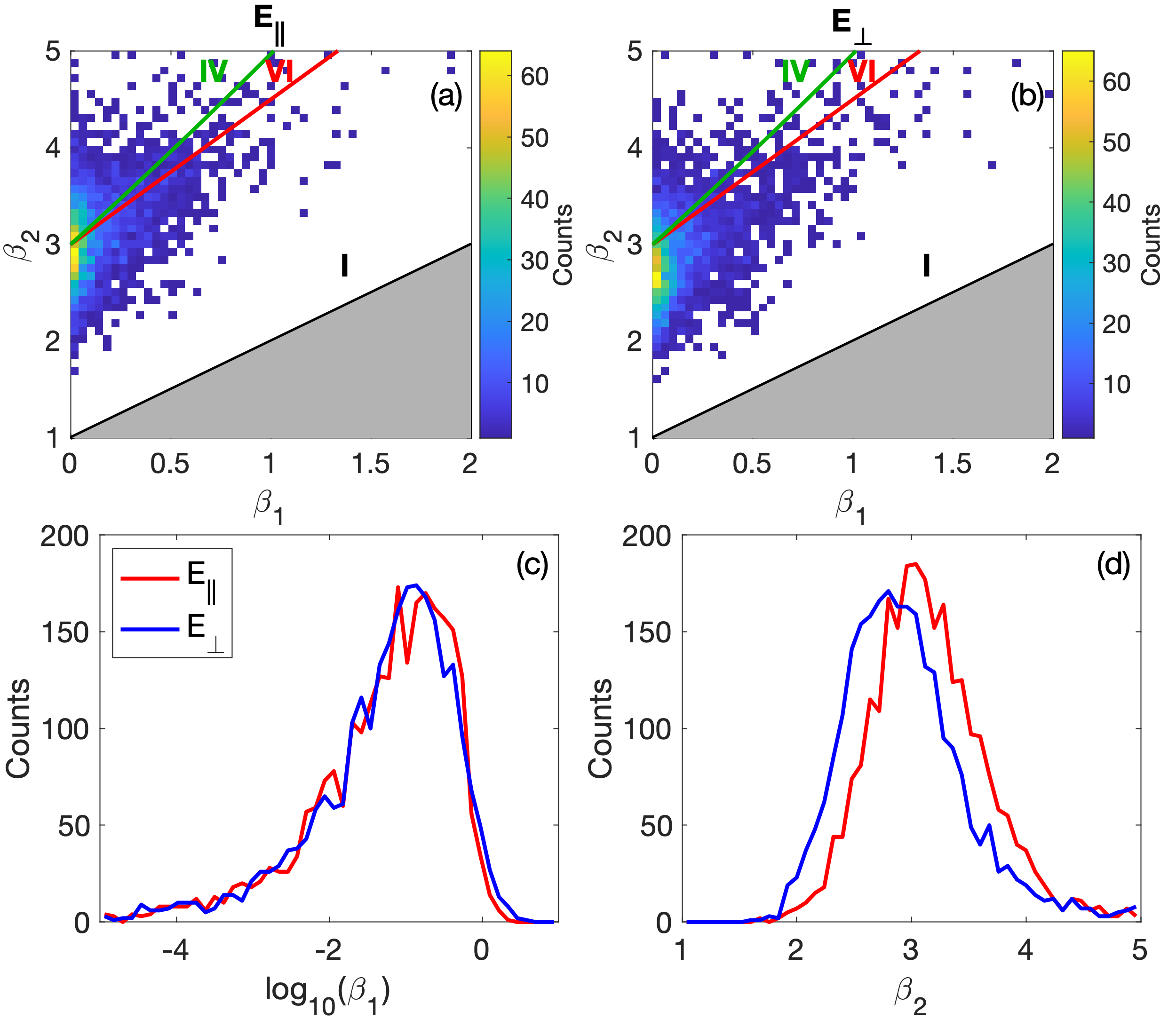}
\caption{Statistics of $\beta_1$ and $\beta_2$ for $E_{\parallel,\mathrm{env}}$ and $E_{\perp,\mathrm{env}}$. (a) and (b) Two-dimensional histograms of $\beta_1$ and $\beta_2$ for $E_{\parallel,\mathrm{env}}$ and $E_{\perp,\mathrm{env}}$, respectively. The gray-shaded region corresponds to $\beta_2 < \beta_1+1$, which is not possible. The red and green lines mark indicate the boundaries between type I and type VI, and type VI and type IV Pearson distributions, respectively. (c) Histograms of $\log{\beta_1}$ for $E_{\parallel,\mathrm{env}}$ (red) and $E_{\perp,\mathrm{env}}$ (blue). (d) Histograms of $\beta_2$ for $E_{\parallel,\mathrm{env}}$ (red) and $E_{\perp,\mathrm{env}}$ (blue). }
\label{beta12statsEpp}
\end{center}
\end{figure}

In Figure \ref{beta12statsEpp} we present the statistical results of $\beta_1$ and $\beta_2$ for $E_{\parallel,env}$ and $E_{\perp,env}$. Figure \ref{beta12statsEpp}a shows that for $E_{\parallel,env}$ are typically characterized by small $\beta_1$ and $\beta_2$ close to $3$, consistent with SGT. We find that for $E_{\parallel,\mathrm{env}}$, $\beta_2$ is on average slightly larger compared with $E_{env}$, with a mean and median of $3.2$ and $3.1$ (Figure \ref{beta12statsEpp}d). Figure \ref{beta12statsEpp}c shows that typically $\beta_1 \ll 1$, with a mean and median of $0.17$ and $0.08$. From the fits to $P(\log{E})$, we find that $49\,\%$ of the snapshots are better fitted by nonlinear SGT than linear SGT based on $\chi_r^2$. We find that $37\,\%$ of the snapshots have $(\beta_1,\beta_2)$ consistent with nonlinear SGT (Figure \ref{sgtnl}). Based on the values of $(\beta_1,\beta_2)$ we find that $62$~\% of the waveforms are type I, $33$~\% are type IV, and $6$~\% are type VI. 

For $E_{\perp,\mathrm{env}}$ we find that $P(\log{E})$ are typically characterized by small $\beta_1$ and $\beta_2 < 3$ (Figure \ref{beta12statsEpp}b). Specifically, $\beta_2$ has a mean and median of $3.0$ and $2.9$. Thus, $\beta_2$ tends to be lower for $E_{\perp,\mathrm{env}}$, compared with $E_{\parallel,env}$ (Figure \ref{beta12statsEpp}d). Figure \ref{beta12statsEpp}c shows that typically $\beta_1 \ll 1$, with a mean and median $0.19$ and $0.08$. We find that $32 \, \%$ of snapshots have $(\beta_1,\beta_2)$, consistent with nonlinear SGT.  From the fits to $P(\log{E})$, $43\,\%$ of the snapshots are better fitted by nonlinear SGT than linear SGT, based on $\chi_r^2$. Based on the values of $(\beta_1,\beta_2)$ we find that $78$~\% of the waveforms are type I, $18$~\% are type IV, and $4$~\% are type VI. This indicates that $E_{\perp,\mathrm{env}}$ and $E_{\parallel,\mathrm{env}}$ differ statistically from each other, which may suggest that the evolution of parallel and perpendicular electric fields differs. 

In summary, we find that the observed $P(\log{E})$ are often characterized by $(\beta_1,\beta_2)$ close to $(0,3)$. In particular, we find that about one third of the waveforms are characterized by $(\beta_1,\beta_2)$ within the range of values predicted by SGT when electrostatic decay occurs. We find that the values of $\beta_2$ on average differ for $E_{\parallel,\mathrm{env}}$ and $E_{\perp,\mathrm{env}}$. The presence of strong $E_{\perp,\mathrm{env}}$, suggestive of Z-mode-like waves, may contribute to the deviation of $E_{\mathrm{env}}$ from the SGT predictions in some cases. Overall, we conclude that the observed Langmuir/Z-mode wave waveforms show good agreement with SGT in the presence of nonlinear processes, suggesting that density fluctuations play an important role in the evolution of the waves. 

\section{Discussion} \label{discussion}
In this section, we discuss the results of the previous section and compare them with previous observations and simulations. We also discuss the role of density perturbations, the role of nonlinear processes associated with Langmuir waves, and future work required to further understand the behavior of Langmuir/Z-mode waves. 

\subsection{Discussion of results and comparisons}
From section \ref{fieldstatssec} we found that many of the observed waveforms are consistent with SGT with and without electrostatic decay. However, we also find that most of the distributions of $\log{E}$ are characterized by $(\beta_1,\beta_2)$ outside the range predicted by SGT, as currently understood. There are several effects that can cause $(\beta_1,\beta_2)$ to deviate from the predictions of linear and nonlinear SGT: 
\begin{enumerate}
\item SGT assumes that the plasma environment is close to marginal stability for Langmuir waves with the gain $G$ undergoing a random walk. A large number of beam interactions are required for the central limit theorem to apply, so that $G$ is normally distributed. For MMS, the orbit typically remains close to the bow shock (Figure \ref{Foreshockposstats}), so in some cases it is possible that the number of fluctuations in growth and damping is small, meaning $G$ may not be normally distributed. Similarly, if the length of snapshots is too short, a normal distribution may not be obtained. For MMS, converting the nominal snapshot duration to distance, assuming convection by the solar wind flow, yields $d \approx 7 \times 10^4\,\lambda_D$. This is much larger than the size of the Langmuir wave clumps and the typical separations between them, making the snapshots well-suited to testing SGT predictions. 

\item The instrumental noise will increase $E$ for very low-amplitude waves, which can modify $P(\log{E})$ at low $\log{E}$. Additionally, the sensitivity of EDP on MMS prevents us from studying $P(\log{E})$ at very low-amplitude thermal levels of $E$. This results in a limit for low values of $\log{E}$ and could potentially affect $(\beta_1,\beta_2)$. In section \ref{fieldstatssec}, we set a threshold on $\mathrm{min}(E_{\mathrm{env}})$ to minimize this effect.

\item In the electron foreshock, the electron beams exciting Langmuir waves are produced by electrons reflected at Earth's bow shock. For typical solar wind conditions, Earth's quasi-perpendicular bow shock is non-stationary and is characterized by ripples along the shock surface at proton kinetic scales \cite[]{lotekar2025}, resulting in a variable local shock-normal angle. Since the electron beam depends strongly on the local shock normal, the electron beam properties can substantially change over the nominal 2~s duration of the snapshots. We note that many of the snapshots analyzed in section \ref{wavepropssec} are characterized by intense Langmuir waves, along with periods of no wave activity above the noise level, suggesting that the local shock-normal angle and the properties of the electrons reflected at the bow shock can vary significantly over durations comparable to the hmfe snapshots. Such snapshots were excluded from the analysis in section \ref{fieldstatssec}. However, strong variations in the reflected electrons over the snapshot time could cause the observed $P(\log{E})$ to deviate from SGT predictions, even when Langmuir waves persist throughout the snapshot. 

\item On MMS, ${\bf E}$ is measured by EDP, which consists of two instruments, namely, the Spin-plane Double Probes (SDP) and the Axial Double Probes (ADP), which have distinct baselines and gains applied during calibration \cite[]{ergun3,lindqvist1}. Due to small uncertainties in the gains between the two instruments, the direction of ${\bf E}$ has some uncertainty. In most cases considered $E_{\parallel} \gg E_{\perp}$, so part of the observed $E_{\perp}$ may result from the uncertainty in relative gains applied to SDP and ADP. This could affect $P(\log{E})$ of $E_{\perp,env}$.  
\end{enumerate}

These effects described above could cause the observed $P(\log{E})$ to deviate from the predictions of linear and nonlinear SGT in some cases. The fact that about one third of the snapshots analyzed in section \ref{sgtEtot} are characterized by $(\beta_1,\beta_2)$ within the range predicted for nonlinear SGT suggests that the Langmuir waves in Earth's electron foreshock are often characterized by stochastic growth due to density perturbations in the solar wind, but with electrostatic decay at high $E$.

The distributions of $\log{E}$ have been calculated from numerical simulations of Langmuir wave evolution in the presence of density fluctuations. \cite{li2006} found good agreement with SGT, using quasi-linear simulations. However, the results from more recent simulations found that $P(\log{E})$ differ from the predictions of SGT \cite[]{voshchepynets2015a,voshchepynets2017,annenkov2025}. For example, \cite{voshchepynets2017} found $P(\log{E})$ was typically characterized by $\beta_2 > 3$ and significant $\beta_1$, corresponding to Pearson type VI and type IV distributions. Similar $P(\log{E})$ were found in \cite{annenkov2025}. These values of $P(\log{E})$ differ significantly from our observations, suggesting further work is needed to reconcile observations with simulations. 

Past observations of Langmuir waves in the electron foreshock and type III source regions have found that the electric field distributions show good agreement with SGT predictions \cite[]{robinson1993a,cairns1997b,cairns1999,sigsbee2004a,sigsbee2004b}. More recent studies have shown that small deviations from the linear SGT prediction \cite[]{musatenko2007,krasnoselskikh2007,vidojevic2011,vidojevic2014}. However, these studies relied on intermittent observations from spectrograms of wave power, or a small number of points from a series of short snapshots, rather than following the rapidly varying $E_{\mathrm{env}}$. These results are thus more similar to those in section \ref{subsecstats}. We note that these studies did not consider the effect of nonlinear processes, which can in part explain the deviation from the prediction of linear SGT. A recent study of Langmuir waves in type III source regions \cite[]{cairns2026} shows good agreement with nonlinear SGT. 

%Statistically, the observed $P(\log{E})$ of $E_{\mathrm{env}}$, $E_{\parallel,\mathrm{env}}$ , $E_{\perp,\mathrm{env}}$ are characterized by $\beta_1 \ll 1$ and $\beta_2 \lesssim 3$. This is in part explained by SGT including electrostatic decay. According to the Pearson classification, $P(\log{E})$ may be modeled by the symmetric beta distribution (Type II distribution). Further work is required to determine if the Pearson system of distribution functions better models the observed $P(\log{E})$ and whether this can shed further light on the underlying processes determining Langmuir wave evolution in the electron foreshock. 

\subsection{Role of density perturbations}
We now discuss the role of density fluctuations in explaining the observed behavior of Langmuir waves. In the presence of density fluctuations, the dispersion relation of electrostatic Langmuir waves in the spacecraft frame with negligible $k_{\perp}$ is given by \cite[]{robinson1992a}
\begin{equation}
\omega_L = \omega_{pe} + \omega_{pe} \frac{\delta n_e}{2 n_e} + \frac{3 v_e^2 k^2}{4 \omega_{pe}} + V_{sw} k \cos{\theta_{kV}},
\label{dispreldn}
\end{equation}
where $v_e = \sqrt{2 k_B T_e/m_e}$ is the electron thermal speed, $V_{sw}$ is the solar wind speed, and $\theta_{kV}$ is the angle between ${\bf k}$ and ${\bf V}_{sw}$. The final term in equation (\ref{dispreldn}) is the Doppler shift and can be positive or negative depending on whether ${\bf k}$ (and the beam velocity) is aligned or anti-aligned with ${\bf V}_{sw}$. Density fluctuations can directly modify the observed $\omega_L$ or can cause $k$ to increase or decrease as the waves propagate across density gradients, leading to growth and damping of the waves. 

By assuming that the changes in $f$ are due to density perturbations and that $k$ is small, such as when prominent Z-mode waves are present, we can estimate the density perturbations as 
\begin{equation}
\frac{\delta n_e}{n_e} \approx \frac{2 (\omega_L - \langle \omega_L \rangle)}{\langle \omega_L \rangle},
\label{dnneq}
\end{equation}
where $\omega_L$ is the instantaneous wave frequency and $\langle \omega_L \rangle$ is the mean frequency. 

\begin{figure*}[htbp!]
\begin{center}
\includegraphics[width=160mm, height=80mm]{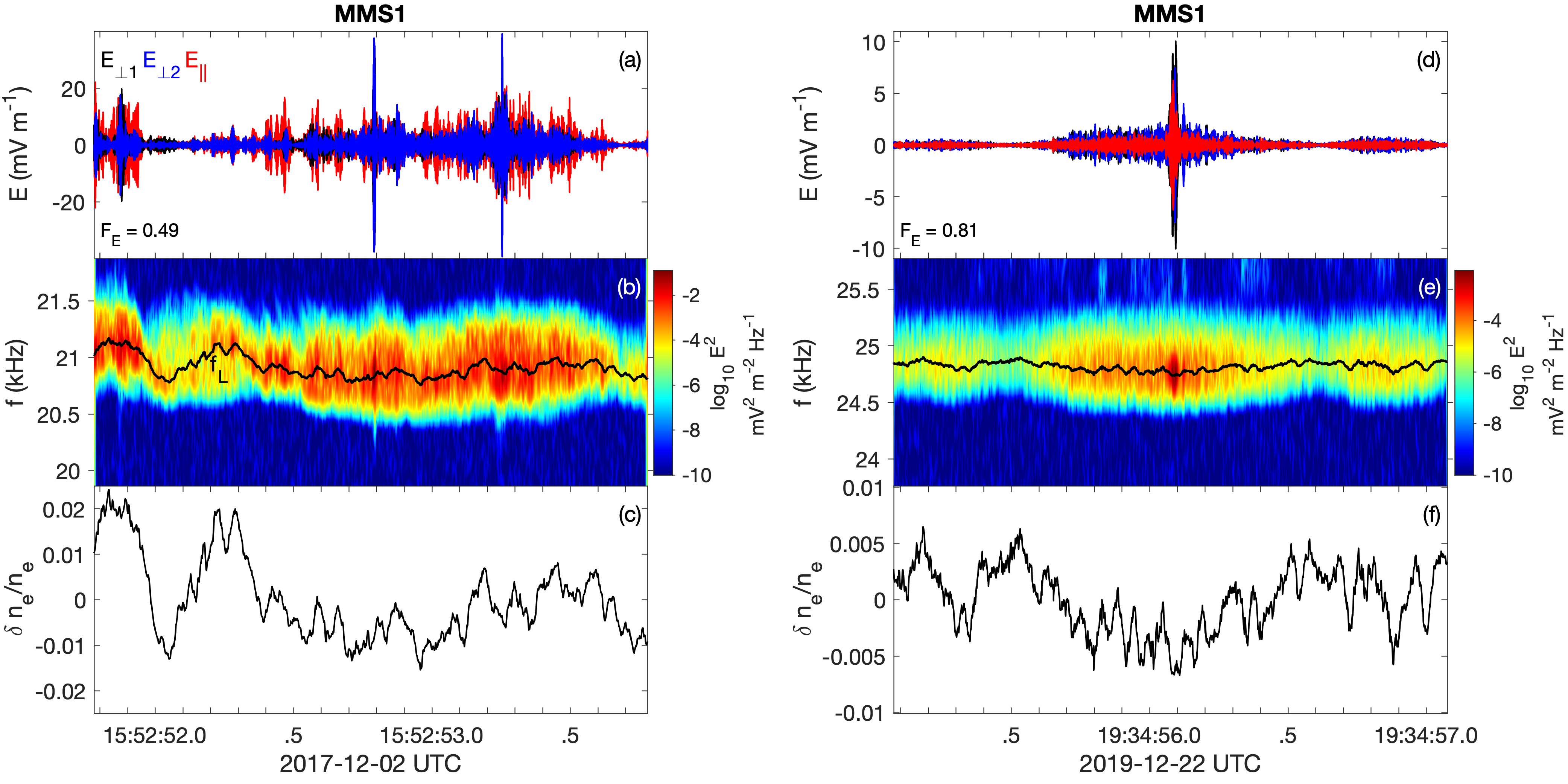}
\caption{Two Examples of Langmuir waves and density perturbations estimated from frequency fluctuations [equation (\ref{dnneq})]. (a)--(c) Langmuir/Z-mode wave observed by MMS1 on 2017 December 02 and Langmuir/Z-mode waves observed by MMS1 on 2019 December 22. (a) and (d) Electric field waveform in field-aligned coordinates $E_{\parallel}$ (red), $E_{\perp1}$ (black), and $E_{\perp2}$ (blue). (b) and (e) Spectrograms of ${\bf E}$. The black lines show the instantaneous Langmuir/Z-mode wave frequency $f_L$. (c) and (f) $\delta n_e/n_e$ estimated from $f_L$.}
\label{dnnefig}
\end{center}
\end{figure*}

Figure \ref{dnnefig} shows two examples of $\delta n_e/n_e$ estimated from the observed waveforms. The waveforms are characterized by strong ${\bf E}_{\perp}$, suggestive of low-$k$ Z-mode waves. The instantaneous angular frequency is calculated using $\omega_L(t) = d \phi(t)/d t$, where $\phi(t)$ is the signal phase calculated using a Hilbert transform of ${\bf E}$. The instantaneous ordinary frequency $f_L(t) = \omega_L(t)/2\pi$, plotted in Figure \ref{dnnefig}, is an average of the three components of ${\bf E}$ weighted by their powers and is smoothed using a moving mean over 2,000 points. 

The first example, Figures \ref{dnnefig}a--\ref{dnnefig}c, shows a highly bursty waveform similar to the previous examples. The spectrogram (Figure \ref{dnnefig}b) shows that frequency variations occur throughout the snapshot. We find that the instantaneous ordinary frequency $f_L$ similarly varies and is typically centered around the peak wave power. Figure \ref{dnnefig}c shows $\delta n_e/n_e$ calculated from equation (\ref{dnneq}). We observe a large oscillation with $\delta n_e/n_e \approx 0.015$ and a period of $T \approx 0.5$~s at the beginning of the snapshot. We also observe smaller scale fluctuations with $\delta n_e/n_e \approx 0.005$ and $T \lesssim 0.1$~s. These periods correspond to length scales of $l \approx 2 \times 10^4 \, \lambda_D$ and $l \lesssim 4 \times 10^3 \, \lambda_D$, respectively. 

Figures \ref{dnnefig}d--\ref{dnnefig}f show a waveform characterized by waves localized near the center of the snapshot and lower amplitude waves extending across the interval. We find that $f_L$ closely traces the peak in wave power (Figure \ref{dnnefig}e). We observe small-amplitude fluctuations in $f_L$ across the interval. 

Figure \ref{dnnefig}f shows similar $\delta n_e/n_e$ to the previous example, although the amplitudes are smaller. We observe large-scale variations with $\delta n_e/n_e \approx 5 \times 10^{-3}$. In this case, the largest-amplitude waves occur where $\delta n_e/n_e$ is minimal, while near the larger-scale maxima in $\delta n_e/n_e$, wave power is reduced. Langmuir waves can become trapped in local small-amplitude density depletions, forming localized eigenmodes \cite[]{ergun2008}. 
We also observe smaller $\delta n_e/n_e \approx 2.5 \times 10^{-3}$ with $T \lesssim 0.1$~s.

We conclude that small-scale density fluctuations are present with amplitudes of $\delta n_e/n_e \lesssim 0.01$ and scales of thousands of $\lambda_D$. These results are consistent with previous estimates of small-scale density perturbations in the solar wind at 1~au \cite[]{celnikier1987,malaspina2010b,malaspina2011}. Such density perturbations can strongly affect the behavior of Langmuir/Z-mode waves. The main effects of small-scale $\delta n_e/n_e$ are: 
\begin{enumerate}
    \item The density perturbations can cause ${\bf k}$ to be shifted in and out of resonance with an electron beam, leading to localized regions of growth and damping, as proposed by SGT. Assuming $f_L$ is conserved in the plasma frame, $k$ will decrease as the waves propagate into higher density regions, and increase in lower density regions. 
    \item Sufficiently large $\delta n_e/n_e$ can cause beam-driven Langmuir waves to partly reflect off density gradients, as well as have their $k$ shifted to very small values, where the wave is Z-mode-like, and can undergo mode conversion to O-mode waves. Assuming $f_L$ is constant in the plasma frame, the condition required for reflection or mode conversion of beam-driven Langmuir waves is 
    \begin{equation}
    \frac{\delta n_e}{n_e} \gtrsim \frac{3}{2} \frac{v_e^2}{v_b^2},
    \end{equation}
    where $v_b$ is the electron beam speed. Assuming a beam speed of $v_b = 1$~keV, $\delta n_e/n_e \gtrsim 0.02$ would be required for reflection or mode conversion, although for faster beams, smaller $\delta n_e/n_e$ will be required. The analysis in Figure \ref{dnnefig} suggests that in some cases Langmuir waves could reflect off density gradients for very fast electron beams, but is not likely for slow beams. However, more detailed analyses are needed. 
    \item The shift in $k$ due to density fluctuations can cause strong amplitude fluctuations. Assuming no growth or damping, the wave energy flux will be conserved along the propagation direction. This corresponds to $E \propto k^{-1/2}$ assuming equation (\ref{dispreldn}) holds, due to changes in group speed as a function of $k$. For Z-mode waves the group velocity can differ significantly and the wave acquires a significant electromagnetic component. For random density fluctuations, where $k$ can reach low values, we predict a power-law distribution at large $E$. From the waveforms analyzed in section \ref{fieldstatssec}, we see very few cases of power-law distributions over a wide range of $E$. We note that ${\bf B}$ is typically oblique to ${\bf V}_{sw}$, so the spacecraft trajectory will be oblique to the typical ${\bf k}$ of Langmuir waves. Therefore, along the spacecraft trajectory we do not necessarily expect wave energy flux to be conserved.
    \item Larger-amplitude $\delta n_e/n_e$ can potentially modify the development of electrostatic decay. For example, simulations have shown that large $\delta n_e/n_e$ can limit the development of electrostatic decay to localized regions \cite[]{krafft2015,krafft2024}, compared with a homogeneous background plasma. 
\end{enumerate}

Further work is needed to understand and quantify the properties of small-scale $\delta n_e/n_e$, such as the typical amplitude of $\delta n_e/n_e$ and whether $\delta n_e/n_e$ develop along ${\bf B}$, and hence the typical direction of ${\bf k}$, or perpendicular to ${\bf k}$. Previous observations show that density turbulence can be anisotropic at small scales \cite[]{celnikier1987,malaspina2010b}, which can be important for interpreting the behavior of Langmuir waves. 

In summary, we conclude that density fluctuations play a fundamental role in the evolution of Langmuir waves in the electron foreshock. We estimate that $\delta n_e/n_e \lesssim 0.01$ at scales below $\sim 10^4 \, \lambda_D$. This is consistent with the results in section (\ref{fieldstatssec}), which show good agreement with SGT. 

\subsection{Role of nonlinear processes}
\begin{figure}[htbp!]
\begin{center}
\includegraphics[width=90mm, height=40mm]{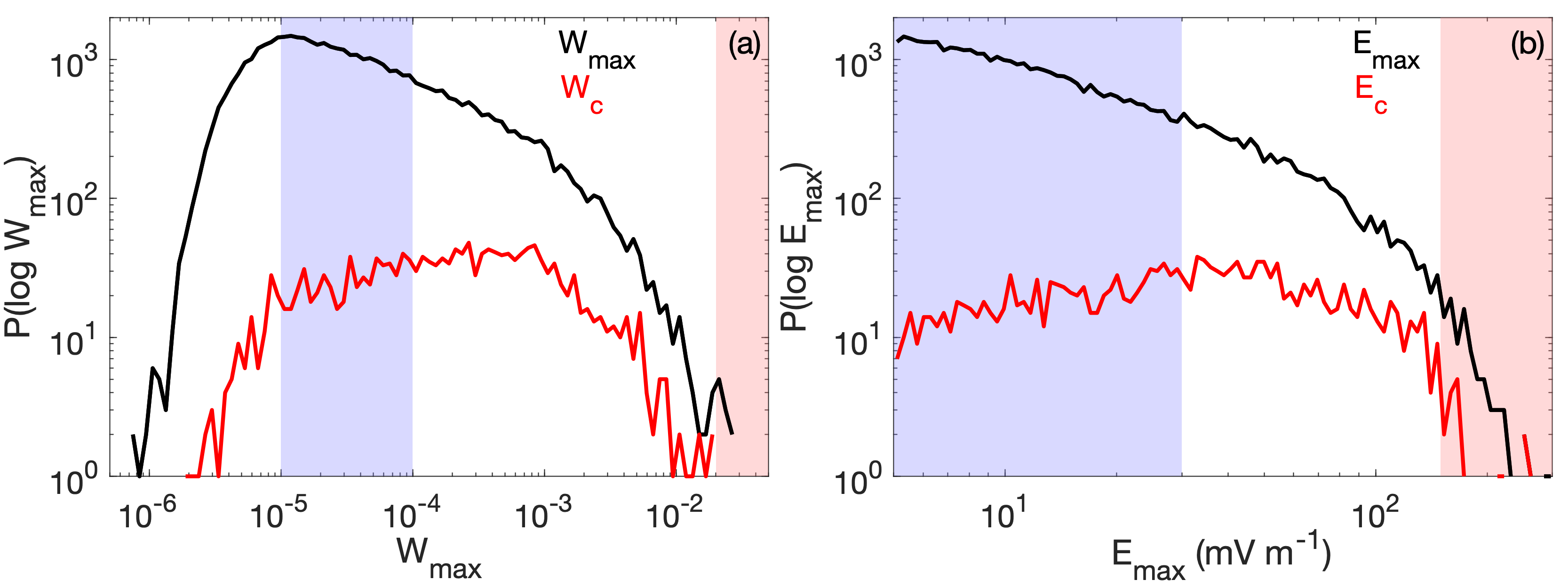}
\caption{Distribution of the peak normalized electric field energy densities $W_{\mathrm{max}}$ and $E_{\mathrm{max}}$. (a) Distributions of $P(\log W_{\mathrm{max}})$ (black) and $P(\log W_c)$ (red). (b) Distributions of $P(\log E_{\mathrm{max}})$ (black) and $P(\log E_c)$ (red). The blue- and red-shaded regions indicate the approximate thresholds for electrostatic decay and wave-packet collapse.}
\label{Endendists}
\end{center}
\end{figure}

We now consider the role of nonlinear processes in determining the behavior of the observed Langmuir waves. In the previous sections, we found evidence that electrostatic decay was occurring, namely, the observation of Langmuir/Z-mode waves with distinct spectral peaks consistent with counter-propagating Langmuir waves and $P(\log{E})$ consistent with electrostatic decay. In the weak turbulence approximation, the threshold for electrostatic decay to proceed is \cite[]{robinson1993b,robinson1995b,graham2013a}
\begin{equation}
W_{\mathrm{max}} \gtrsim \frac{3}{16 \sqrt{2} \pi} \sqrt{\frac{m_p}{m_e}} \frac{\gamma_L'}{\omega_{pe}} \frac{\Delta v_b}{v_b} \frac{v_e}{v_b},
\label{ESdecayeq}
\end{equation}
where $W_{\mathrm{max}} = \epsilon_0 E_{\mathrm{max}}^2/4 n_e k_B T_e$ is the maximum electric field energy density normalized to the electron thermal energy density, $m_p/m_e$ is the proton to electron mass ratio, $\gamma_L'$ is the damping rate of the backscattered Langmuir waves, $v_b$ is the electron beam speed, and $\Delta v_b$ is the spread in the electron beam speed. Assuming a nominal solar wind electron temperature of $T_e = 12$~eV \cite[]{newbury1998}, $\Delta v_b \approx v_e$, $v_b$ corresponding to $\sim 1$~keV, and $\gamma_L' \approx 10^{-3} \, \omega_{pe}$ we estimate that $W_{\mathrm{max}} \sim 10^{-5}$--$10^{-4}$ is required for electrostatic decay to proceed. This corresponds to $E_{\mathrm{max}} \gtrsim 5$~mV~m$^{-1}$--$30$~mV~m$^{-1}$, depending on plasma conditions. Figures \ref{Endendists}a and \ref{Endendists}b show the distributions of $W_{\mathrm{max}}$ and $E_{\mathrm{max}}$ for the electron foreshock Langmuir waves (section \ref{wavepropssec}), along with approximate threshold values. We find that $P(\log W_{\mathrm{max}}) \propto W_{\mathrm{max}}^{-0.45}$ for $10^{-5 }\lesssim W_{\mathrm{max}} \lesssim 10^{-3}$, consistent with Figure \ref{probdist1}. The sharp drop in $P(\log W_{\mathrm{max}})$ for $W_{\mathrm{max}} \lesssim 10^{-5}$ is due to the selection criteria of $E_{\mathrm{max}} > 5$~mV~m$^{-1}$. We find that the electrostatic decay threshold is satisfied in most cases. We also plot the distributions of the normalized energy densities $W_c$ corresponding to $E_c$, and $E_c$ for waveforms with $P(\log E)$ better fitted by nonlinear SGT in Figures \ref{Endendists}a and \ref{Endendists}b (section \ref{sgtEtot}). We find that $W_c$ and $E_c$ have values comparable to, but typically larger than the predicted threshold for electrostatic decay. We conclude that the amplitude of the observed waves is consistent with  electrostatic decay frequently occurring in the electron foreshock. 

%We note that strong $\delta n_e/n_e$ can potentially suppress electrostatic decay. For example, \cite{krafft2014a} found that 

We now discuss the role of modulational instabilities and wave-packet collapse in the electron foreshock. Modulational instabilities are four-wave interactions, where relatively uniform low-$k$ Langmuir waves are pumped to higher $k$, modulating the electric field envelope \cite[]{zakharov1985,robinson1997,cairns1998a}. We find that modulational instabilities are unlikely to be relevant in any of the observed Langmuir wave event for the following reasons: 
\begin{enumerate}
    \item Modulational instabilities are predicted to occur for $v_b \gtrsim 3 v_e \sqrt{m_p/(2 m_e)}$, corresponding to $v_b/c \gtrsim 0.6$. It is unlikely that such large $v_b$ occur and excite Langmuir waves in Earth's electron foreshock \cite[]{cairns1998a}. The fact that most Langmuir/Z-mode waveforms are characterized by $F_E \sim 0$ suggests a typical beam speed of $v_b/c \lesssim 0.08$ based on the results in \cite{malaspina2011} and \cite{graham2014b}, where large $F_E$ waveforms were associated with $v_b/c \gtrsim 0.08$. 
    \item The threshold for modulational instabilities is generally based on the unmagnetized approximation, where three-wave decay is prohibited at very small $k$. However, in weakly magnetized plasmas the decay of Langmuir-like waves to Z-mode-like waves can proceed \cite[]{layden2013,cairns2018,polanco2025} and may be favored over modulational instabilities. 
    \item Modulational instabilities require extremely narrow bandwidth waves to occur \cite[]{cairns1998a}, or equivalently, very uniform electric fields. The results in Figure \ref{probdist1}c and section \ref{fieldstatssec} show that the waveforms are highly bursty, meaning modulation instabilities are unlikely to develop. 
\end{enumerate}

Wave-packet collapse occurs when Langmuir waves become trapped in density cavities, intensify, and collapse to smaller scales where they are dissipated. Wave-packet collapse can result from modulation instabilities, but can occur via nucleation in the solar wind, where background density perturbations are present, allowing Langmuir waves to refract into them \cite[]{robinson1997}. In this situation, Langmuir waves refract and amplify in localized regions of reduced density, where the Langmuir wave reflective index increases \cite[]{zakharov1985,robinson1997}. If the localized $W_{\mathrm{max}}$ is sufficiently large, localized Langmuir wave packets will undergo collapse, where $W_{\mathrm{max}}$ increases and their characteristic scale decrease until they are damped at small scales of $\sim 20 \lambda_D$ \cite[]{robinson1991}. 
The threshold for wave-packet collapse can be expressed as \cite[]{robinson1990a,graham2012a}
\begin{equation}
W_{\mathrm{max}}(l/\lambda_D)^2 \gtrsim 200,
\label{wpthres}
\end{equation}
where $l$ is the characteristic width of the wave packet. Previous observations of isolated localized Langmuir waves found characteristic sizes of $l \sim 100 \, \lambda_D$ \cite[]{nulsen2007,graham2012a,graham2014b} in planetary foreshocks and type III source regions, although such wave packets did not satisfy the collapse threshold. For $l = 100~\lambda_D$ the threshold condition is $W_{\mathrm{max}} \gtrsim 0.02$, while for $l = 20~\lambda_D$, $W_{\mathrm{max}} \gtrsim 0.5$ is expected. Figure \ref{Endendists}a shows that only $0.02$~\% of the waveforms satisfy $W_{\mathrm{max}} > 0.02$ and no waveforms reach $W_{\mathrm{max}} \sim 0.5$. Therefore, wave-packet collapse is unlikely to occur in Earth's electron foreshock and cannot account for the observed behavior of the Langmuir waves. 

In summary, we find that electrostatic decay is likely to occur in Earth's electron foreshock based on the observed wave amplitudes. However, modulational instabilities and wave-packet collapse are extremely rare or do not occur, and thus cannot explain the typical behavior of the observed waveforms. We conclude that the observed waveforms, such as the highly variable and rapid changes in amplitude, as well as the observation of multiple spectral peaks, result from perturbations in the medium and electrostatic decay.  

\section{Conclusions} \label{conclusions}
In this paper, we have presented a statistical overview of plasma frequency waves observed in the electron foreshock by the four MMS spacecraft. We have focused on the statistical properties of the waves, where they occur in relation to the electron foreshock, and we have presented detailed field statistics of the waveforms of Langmuir/Z-mode waves to determine their evolution. The results apply to type II and III source regions where similar Langmuir/Z-mode waves are also observed. 

The key results are: 
\begin{itemize}
\item Large-amplitude Langmuir/Z-mode waves and beam-mode waves are common in Earth's electron foreshock. The waves can reach maximum amplitudes of several hundred mV~m$^{-1}$. The long snapshots of the waveforms show that the waveforms are characterized by bursty fluctuations, with wave amplitudes varying over time scales of a few ms. 
\item The largest amplitude waves develop close to the boundary between the electron foreshock and solar wind, where the fastest electron beams are expected to occur. Statistically, the amplitude of Langmuir waves decreases with distance downstream of the edge of the electron foreshock. 
\item Field statistics of the total, parallel, and perpendicular components of the electric field exhibit close to log-normal distributions, consistent with stochastic growth and damping of the fields. This suggests that small-scale density fluctuations play an important role in the evolution of Langmuir/Z-mode waves. Deviations in observed distributions from a log-normal distribution can in part be explained by electrostatic decay occurring when the waves reach large amplitudes. 
\end{itemize}

\begin{acknowledgements}
We thank the MMS team for data access and support. DBG and IHC acknowledge support from the ISSI team, \emph{Beam Plasma Interaction in the Solar Wind and the Generation of Type III Radio Bursts}. This work was supported by the Swedish National Space Agency (SNSA), grant 128/17, and the Knut and Alice Wallenberg Foundation (Dnr. 2022.0087). 
\\
MMS data are available at \url{https://lasp.colorado.edu/mms/sdc/public} and \url{https://spdf.gsfc.nasa.gov/pub/data/mms/}. For MMS1, we use burst mode electric field data from EDP \cite[]{mms1edphmfe}, burst mode magnetic field data from FGM \cite[]{mms1fgmbrst}, and burst mode electron and ion moments from FPI \cite[]{mms1fpidesmomsbrst,mms1fpidismomsbrst}. Equivalent data products are used for the other three spacecraft. The data analysis was performed using the irfu-matlab software package \cite[]{khotyaintsev2024soft}. The SDP region calibration files used to identify solar wind intervals are available as part of the irfu-matlab software package. The routines to reproduce the plots and the statistical data used in this study are available at \url{https://doi.org/10.5281/zenodo.20559084} \cite[]{graham2026}.
\end{acknowledgements}

\bibliographystyle{aa}
%\bibliography{magrecpapers}

\begin{appendix}
\section{Comparison of SGT distributions with Pearson distributions} \label{app1}
In this Appendix, we compare the fits of linear SGT and the Pearson system of distribution functions with the observed probability distributions $P(\log{E})$. Specifically, we compare the obtained $\chi_r^2$ to quantify the goodness of fit. We rescale the $\chi^2$ and $\chi_r^2$ calculated from equation (\ref{chi2eq}) by multiplying by $n \cdot \Delta \log{E}$, which is nominally $\approx 10^4$, so that the calculated $\chi^2$ matches the standard Pearson $\chi^2$ test statistic \cite[]{pearson1900}. 

\begin{figure*}[htbp!]
\begin{center}
\includegraphics[width=150mm, height=45mm]{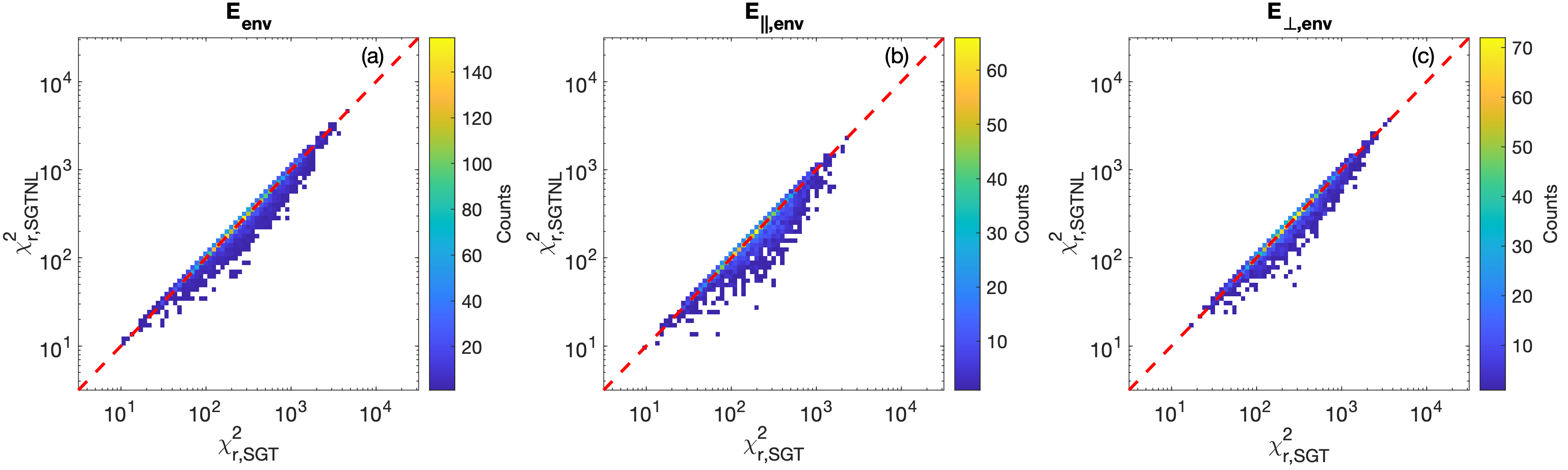}
\caption{Two-dimensional histograms showing the comparison of $\chi_r^2$ calculated from fits of linear SGT [equation (\ref{sgteqlin})] and nonlinear SGT [equation (\ref{sgteqnl})] to the observed $P(\log{E})$ for (a) $E_{env}$, (b) $E_{\parallel,env}$, and (c) $E_{\perp,env}$.}
\label{chi2rSGTNL}
\end{center}
\end{figure*}

We first compare $\chi_r^2$ obtained from the best fits of linear SGT [equation (\ref{sgteqlin})] and nonlinear SGT [equation (\ref{sgteqnl})] to $P(\log{E})$. Figure \ref{chi2rSGTNL} shows the $\chi_r^2$ calculated from the best fit of nonlinear SGT [equation (\ref{sgteqnl})] versus the best fit of linear SGT [equation (\ref{sgteqlin})] to the observed $P(\log{E})$. Figure \ref{chi2rSGTNL}a shows the $\chi_r^2$ comparison for the 5,129 snapshots investigated in section \ref{sgtEtot}. In most cases, we find that the $\chi_r^2$ are comparable, corresponding to approximately equal $\chi^2$ for linear and nonlinear SGT. Some snapshots are characterized by significantly reduced $\chi_r^2$ for nonlinear SGT. Since nonlinear SGT includes one additional free parameter, $\log{E_c}$, the comparable $\chi^2$ results in a smaller $\chi_r^2$ for linear SGT in approximately two-thirds of the snapshots.

Figures \ref{chi2rSGTNL}b and \ref{chi2rSGTNL}c show the same comparison for $P(\log{E})$ calculated from $E_{\parallel,\mathrm{env}}$ and $E_{\perp,\mathrm{env}}$ for the snapshots investigated in section \ref{sgtEparperp}. Overall, we find the same behavior of $\chi_r^2$ as in Figure \ref{chi2rSGTNL}a, namely, most snapshots are characterized by comparable $\chi_r^2$, with a smaller fraction of snapshots where nonlinear SGT provides significantly better fits. 

For the comparison between linear SGT and the Pearson distributions, we consider two cases: 
\begin{enumerate}
\item We calculate $\chi_r^2$ from the observed $P(\log{E})$ compared with the linear SGT prediction [equation (\ref{sgteqlin})] and the general Pearson distribution function [equation (\ref{pearsoneq})] computed from the moments of the observed distribution; namely, $\mu$ and $\sigma$ for linear SGT, and $\mu$, $\sigma$, skewness $\mu_3/\mu_2^{3/2}$, and $\beta_2$ for the Pearson distributions. 
\item We fit the linear SGT prediction and Pearson distribution function to the observed $P(\log{E})$ to minimize $\chi^2$. For fitted distributions, we include an amplitude coefficient in the fitting routine when finding the best fit. 
\end{enumerate}

\begin{figure}[htbp!]
\begin{center}
\includegraphics[width=90mm, height=75mm]{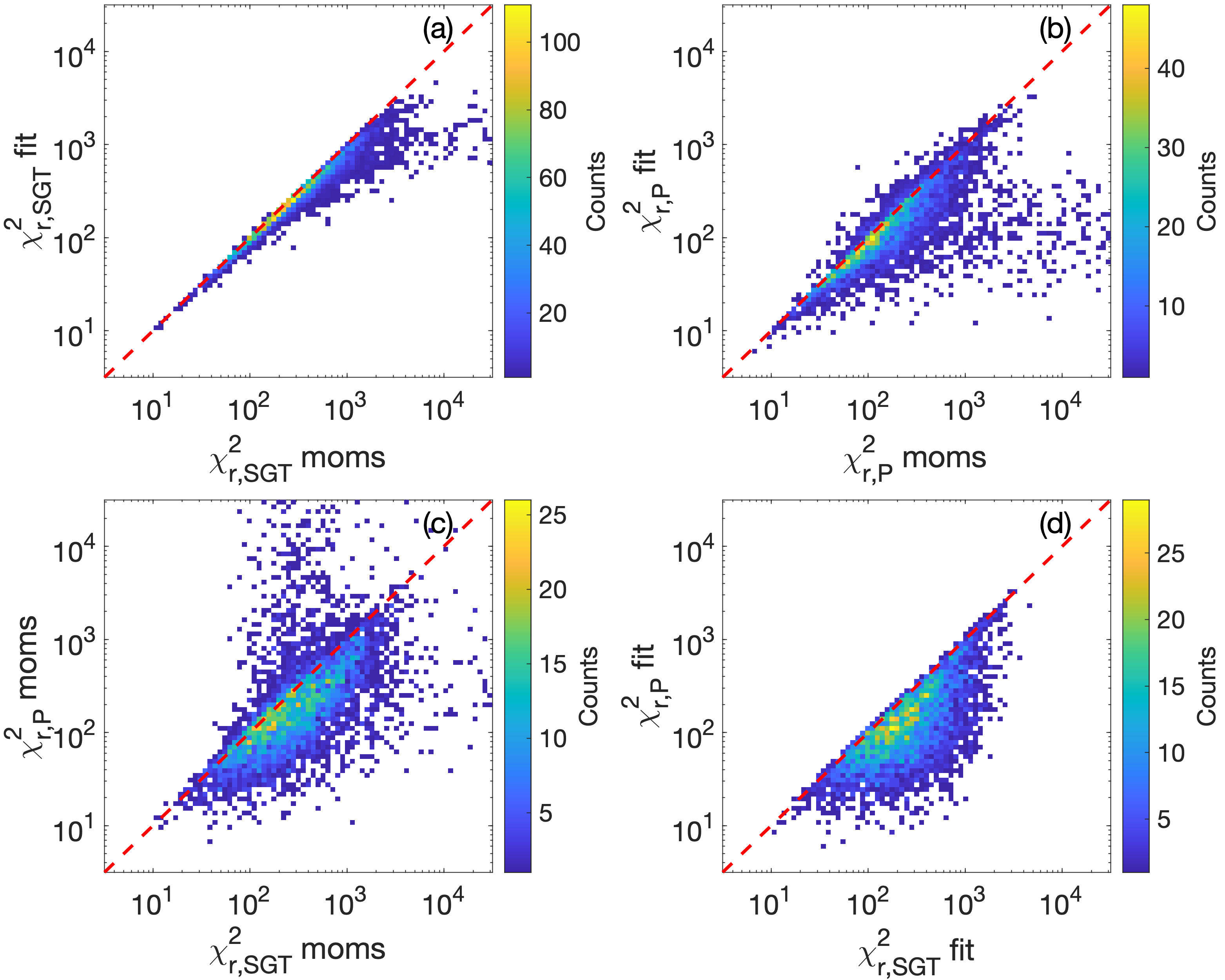}
\caption{Two-dimensional histograms showing the comparison of $\chi_r^2$ calculated from fits of linear SGT and the general Pearson distribution to the observed $P(\log{E})$ for $E_{env}$. (a) Histogram of $\chi_r^2$ calculated from the moments versus $\chi_r^2$ calculated from the fit of SGT to $P(\log{E})$. (b) Histogram of $\chi_r^2$ calculated from the moments versus $\chi_r^2$ calculated from the fit of the Pearson distribution to $P(\log{E})$. (c) Histogram of $\chi_r^2$ computed from the moments for the fit of linear SGT versus the Pearson distribution. (d) Histogram of $\chi_r^2$ of the fit of linear SGT versus the fit of the Pearson distribution to $P(\log{E})$. }
\label{chi2rEtotfig}
\end{center}
\end{figure}

The results of this analysis for $E_{\mathrm{env}}$ of the total electric are shown in Figure \ref{chi2rEtotfig}. We use the 5,129 snapshots analyzed in section \ref{sgtEtot}. In Figure \ref{chi2rEtotfig}a we plot the two-dimensional histogram of $\chi_r^2$ calculated from the fits to the linear SGT prediction from the moments versus the fitting to SGT to minimize $\chi^2$. We find that $\chi_r^2$ is typically comparable in both cases. We find similar results when we compare the $\chi_r^2$ computed from the moments of the Pearson distribution versus the fitting of the Pearson distribution to $P(\log{E})$ [Figure \ref{chi2rEtotfig}b]. The cases where $\chi_r^2$ are smaller when calculated from the moments compared with from the fits minimizing $\chi^2$ primarily result from cases where the calculated model distribution predicts $P_{model} = 0$; these points are neglected when calculating $\chi$ and $\chi_r$. The cases where large values of $\chi_r^2$ are calculated from the moments typically result from large deviations in the observed and modeled $P(\log{E})$ for the largest or smallest $E_{env}$. 

In Figure \ref{chi2rEtotfig}c we plot the histogram of $\chi_r^2$ calculated from the observed $P(\log{E})$ for the linear SGT prediction and the Pearson distribution based on the observed moments. We find that in general $\chi_r^2$ is reduced for the Pearson distribution, although for most snapshots $\chi_r^2$ is comparable, suggesting that Pearson distributions do not typically provide a substantially better fit to the data. Figure \ref{chi2rEtotfig}d shows the histogram of $\chi_r^2$ for the best fit of linear SGT versus the fit of Pearson distributions to $P(\log{E})$. We find that $\chi_r^2$ is reduced for the Pearson distribution fits, since the Pearson distribution accommodates a wide range of distributions, which incorporate two additional free parameters. We find that the median ratio between $\chi_r^2$ for linear SGT and the Pearson distributions is $1.8$. Therefore, the Pearson system of probability distribution functions provides better fits to the observed distributions, although the improvement is relatively minor compared with the prediction from SGT. We conclude that SGT likely applies for many of the observed waveforms, with deviations arising from the effect of nonlinear processes, the central limit theorem not strictly applying, or the contribution of low-$k$ Z-mode waves, which may modify the observed $P(\log{E})$. 

From Figure \ref{chi2rEtotfig}b we find that the fitted Pearson distributions can differ from those predicted from the moments of $P(\log{E})$. Thus, the type of distribution can change between the two cases. From section \ref{sgtEtot} we found, based on $(\beta_1,\beta_2)$, that 75~\% of the snapshots are type I, 22~\% are type IV, and 4~\% are type VI. From $(\beta_1,\beta_2)$ obtained from the best fits of the Pearson distribution to $P(\log{E})$, we find that 38~\% of the snapshots are type I, 59~\% are type IV, and 2~\% are type VI. The fitted distributions have median $\beta_1 = 0.1$ and $\beta_2 = 3.1$, compared with $\beta_1 = 0.07$ and $\beta_2 = 2.9$ calculated from moments of the observed $P(\log{E})$. In both cases $\beta_1 \ll 1$, meaning the increase in median $\beta_2$ for the fitted distributions results in a change in the proportion of distribution types. The fact that the typical values of $\beta_1$ and $\beta_2$ calculated from the moments of $P(\log{E})$ and the fits are often within the range of values predicted by equation (\ref{sgteqnl}) suggests that SGT, along with electrostatic decay, often explains the behavior of the observed waveforms. 

\begin{figure}[htbp!]
\begin{center}
\includegraphics[width=90mm, height=75mm]{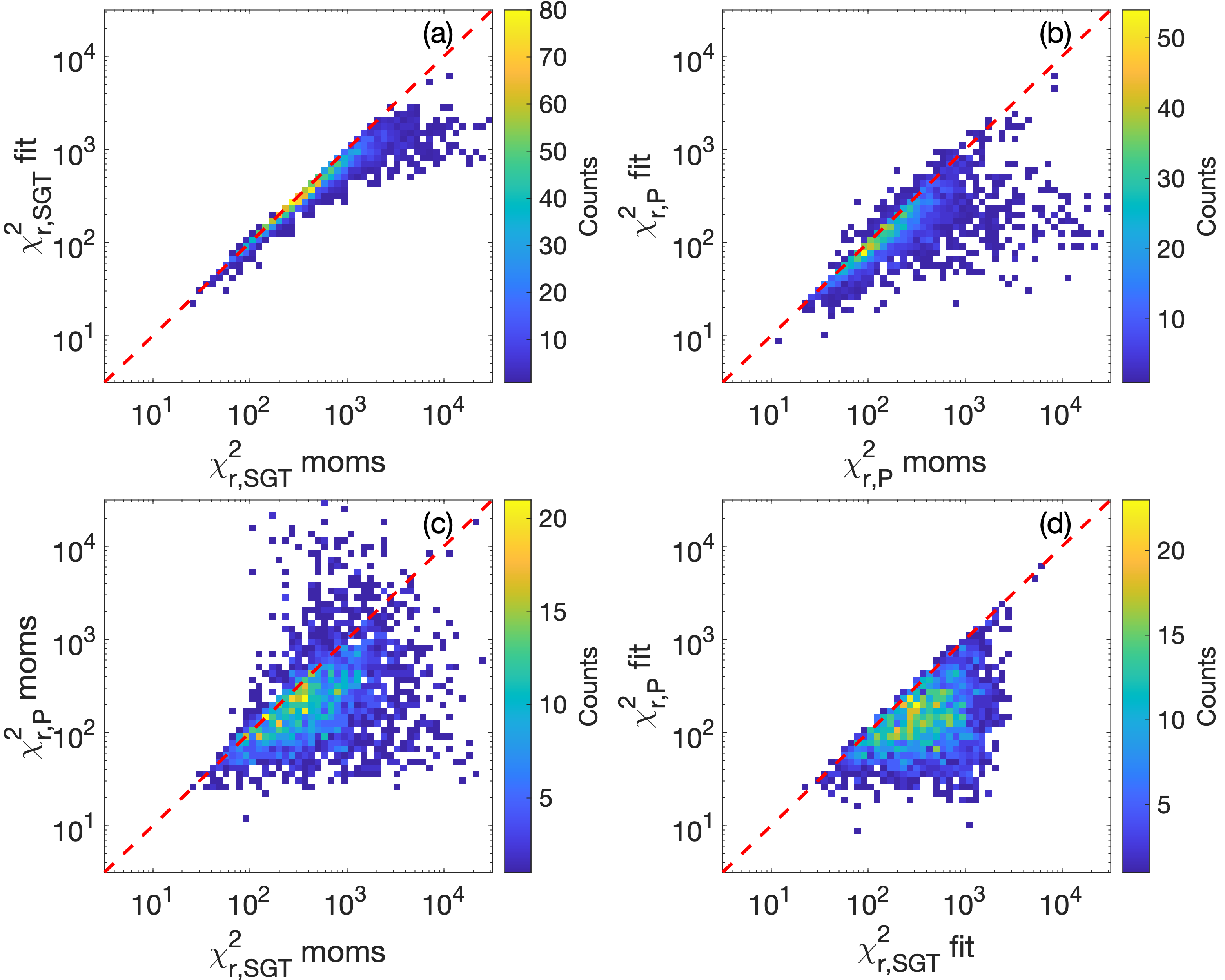}
\caption{Two-dimensional histograms showing the comparison of $\chi_r^2$ calculated from fits of linear SGT and the general Pearson distribution to the observed $P(\log{E})$ for $E_{\parallel,env}$. (a) Histogram of $\chi_r^2$ calculated from the moments versus $\chi_r^2$ calculated from the fit of SGT to $P(\log{E})$. (b) Histogram of $\chi_r^2$ calculated from the moments versus $\chi_r^2$ calculated from the fit of the Pearson distribution to $P(\log{E})$. (c) Histogram of $\chi_r^2$ computed from the moments for the fit of linear SGT versus the Pearson distribution. (d) Histogram of $\chi_r^2$ of the fit of linear SGT versus the fit of the Pearson distribution to $P(\log{E})$. }
\label{chi2rEparfig}
\end{center}
\end{figure}

\begin{figure}[htbp!]
\begin{center}
\includegraphics[width=90mm, height=75mm]{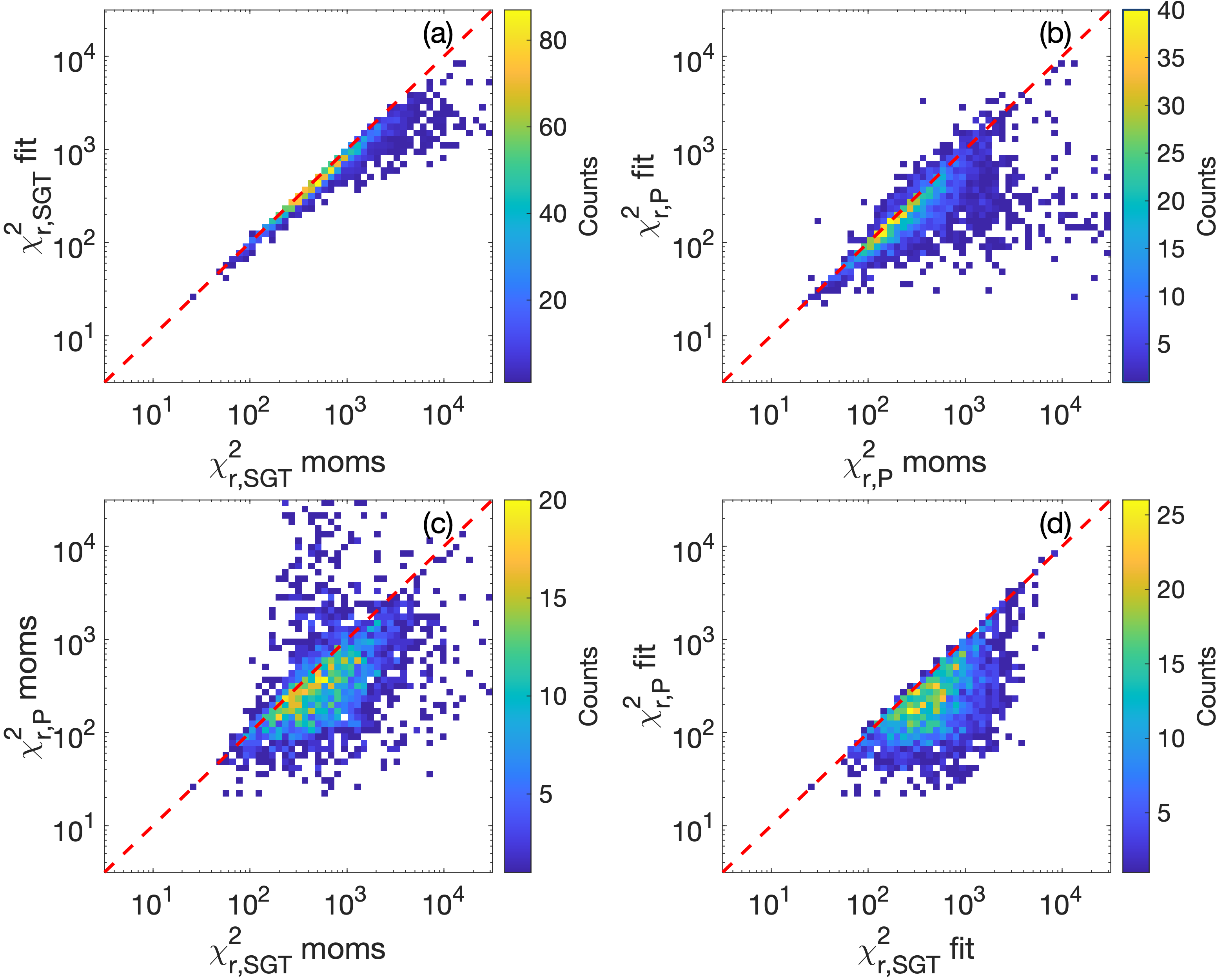}
\caption{Two-dimensional histograms showing the comparison of $\chi_r^2$ calculated from fits of linear SGT and the general Pearson distribution to the observed $P(\log{E})$ for $E_{\perp,env}$. (a) Histogram of $\chi_r^2$ calculated from the moments versus $\chi_r^2$ calculated from the fit of SGT to $P(\log{E})$. (b) Histogram of $\chi_r^2$ calculated from the moments versus $\chi_r^2$ calculated from the fit of the Pearson distribution to $P(\log{E})$. (c) Histogram of $\chi_r^2$ computed from the moments for the fit of linear SGT versus the Pearson distribution. (d) Histogram of $\chi_r^2$ of the fit of linear SGT versus the fit of the Pearson distribution to $P(\log{E})$. }
\label{chi2rEperpfig}
\end{center}
\end{figure}

In Figures \ref{chi2rEparfig} and \ref{chi2rEperpfig} we present the same results as in Figure \ref{chi2rEtotfig} for $P(\log{E})$ calculated from the $E_{\parallel,\mathrm{env}}$ and $E_{\perp,\mathrm{env}}$ for the 2,544 snapshots analyzed in section \ref{sgtEparperp}. We find that for both $E_{\parallel,\mathrm{env}}$ and $E_{\perp,\mathrm{env}}$ the results are similar to those in Figure \ref{chi2rEtotfig}, namely, $\chi_r^2$ are similar for linear SGT and the Pearson distributions calculated from the moments (Figures \ref{chi2rEparfig}a--\ref{chi2rEparfig}b and \ref{chi2rEperpfig}a--\ref{chi2rEperpfig}b). Similarly, the $\chi_r^2$ calculated from the moments for linear SGT are typically larger than, but comparable to, the $\chi_r^2$ from the moments for the Pearson distribution (Figures \ref{chi2rEparfig}c and \ref{chi2rEperpfig}c). The same statistical result is found for the comparison of the fits of linear SGT and the Pearson distribution to the data (Figures \ref{chi2rEparfig}d and \ref{chi2rEperpfig}d).

Finally, we note that the calculated $\chi_r^2$ statistics are characterized by $\chi^2 \gg \nu$ for fits of the linear and nonlinear SGT predictions, as well as the Pearson system of distribution functions, to the observed $P(\log{E})$. Thus, the calculated $\chi^2$ generally correspond to points in the far tail of the $\chi^2$ distribution and are thus unlikely to pass the standard Pearson $\chi^2$ test for goodness-of-fit \cite[]{pearson1900}. We attribute this to the fact that we calculate $\log{E}$ from the continuous waveform of ${\bf E}$, where $E_{\mathrm{env}}$ evolves slowly compared with the sampling rate. This results in an effective over-sampling of $n$, increasing $\chi^2$. This is in contrast to previous observations, which relied on much more sparsely sampled wave amplitudes or powers. We conclude that the standard Pearson $\chi^2$ test for goodness-of-fit calculated from continuous waveform data is too strict when comparing the observed $P(\log{E})$ with predictions of SGT or the Pearson system of distribution functions.

\end{appendix}

\end{document}